\documentclass[10pt,twocolumn]{article}

\usepackage[margin=1in]{geometry}

\usepackage{amssymb}
\usepackage{amsmath}
\usepackage{amsthm}

\usepackage{booktabs}
\usepackage{multirow}

\usepackage{algorithm}
\usepackage{algpseudocode}

\usepackage{graphicx}
\usepackage{subcaption}
\usepackage{tikz}
\usetikzlibrary{arrows.meta,positioning}

\usepackage{url}

\usepackage{fvextra}

\usepackage{placeins}

\usepackage{hyperref}

\usepackage{authblk}

\newif\ifblind
\ifdefined\BLINDMODE
\blindtrue
\else
\blindfalse
\fi

\begin{document}
\flushbottom

\title{GenWorld: Empirically Grounded Urban Simulation Infrastructure for Scalable LLM-Agent Studies}

\ifblind
\author{Anonymous Author(s)}
\date{}
\else
\author[1]{Gen Li}
\author[1,2]{Jieyuan Lan}
\author[3]{Pengcheng Xu}
\author[4]{Zongyuan Wu}
\author[1]{Masaki Ogura}
\author[1]{Tao Feng\thanks{Corresponding author: taofeng@hiroshima-u.ac.jp}}

\affil[1]{Graduate School of Advanced Science and Engineering, Hiroshima University, Higashi-Hiroshima, Japan}
\affil[2]{Jiangxi Polytechnic University, Jiangxi, China}
\affil[3]{College of Urban Development and Modern Transportation, Xi'an University of Architecture and Technology, Xi'an, China}
\affil[4]{North China University of Water Resources and Electric Power, China}

\date{\today}
\fi

\maketitle

\begin{figure*}[t]
\centering
\includegraphics[width=\textwidth]{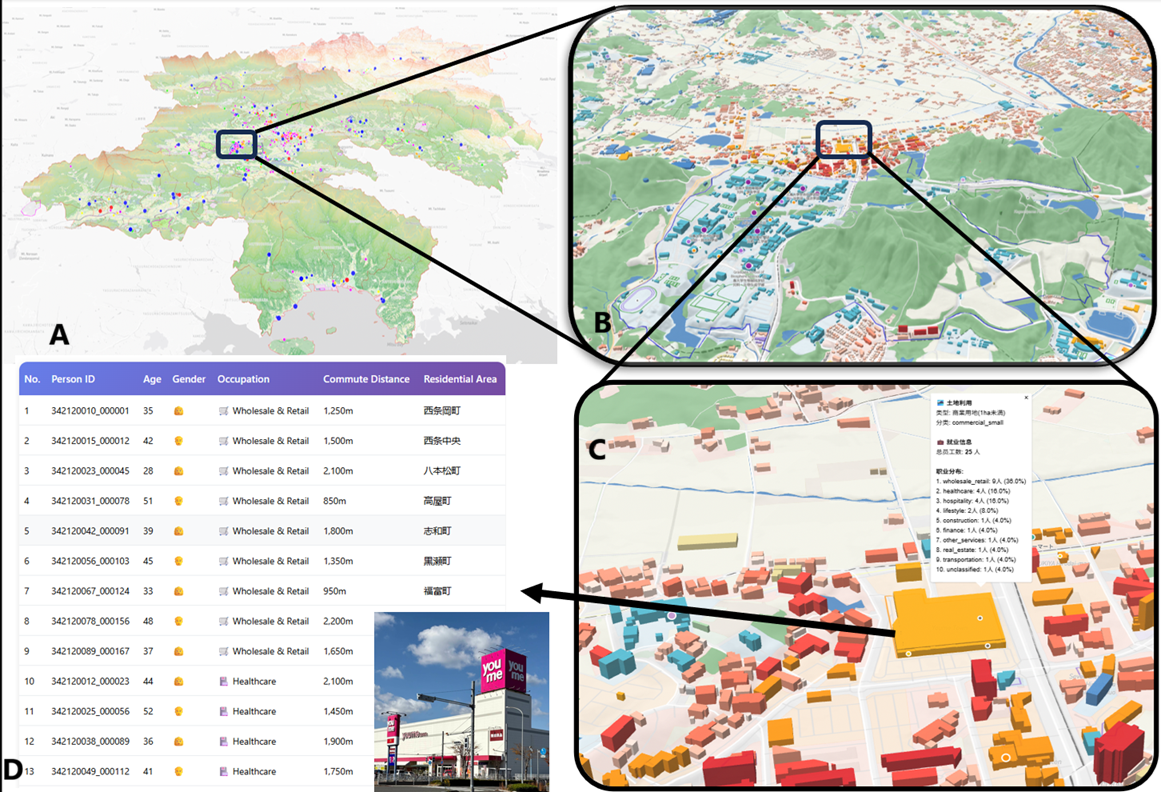}
\caption{Multi-scale spatial granularity of \textbf{GenWorld}'s empirically grounded urban world in Higashihiroshima, Hiroshima, Japan. \textbf{(A)} City-level view showing 196,608 individuals distributed across georeferenced buildings, validated against census data. \textbf{(B)} District-level view near Hiroshima University, revealing diverse building types (residential, commercial, educational) with topographic context and elevation data. \textbf{(C)} Building-level view of a Youme Town supermarket area with 47 employees spatially assigned to a corresponding commercial land-use parcel; residential buildings are rendered in red with color intensity proportional to resident counts (darker indicates more residents). \textbf{(D)} Individual-level details showing employee household origins, occupations, commuting distances, and residential neighborhoods (cho/town). This fine-grained spatial allocation supports environment-aware agent simulation, which is typically difficult to capture in TAZ-based or POI-list approaches.}
\label{fig:platform_overview}
\end{figure*}

\begin{abstract}
LLM-agent simulation faces a joint grounding and scaling problem: agents should act in environments that reflect real urban constraints, yet direct online LLM calls for city-scale populations are computationally prohibitive. We present \textbf{GenWorld}, an empirically grounded urban simulation infrastructure that combines a building-level synthetic city, a structured agent--environment interface, and offline compilation of LLM-derived decision signals into lookup policies for scalable rollout. In a reference instantiation for Higashihiroshima, Japan, GenWorld grounds 196,608 synthetic residents in census and geospatial data, validates demographic consistency against census tabulations, and uses YJMob100K mobile-phone data as a commuting-distance diagnostic. We demonstrate the infrastructure through three reproducible cases: a full-city weekday rollout, a weekday--weekend behavioral contrast, and a warning-response perturbation with auditable replanning traces. These cases support GenWorld as a reproducible platform for grounded and scalable LLM-agent studies, while calibrated forecasting for traffic, evacuation, or policy outcomes remains future work.\ifblind\else Project page: \url{https://genworld1993.netlify.app/}.\fi
\end{abstract}


\section{Introduction}

\subsection{Grounded and Scalable LLM-Agent Simulation}

Large Language Models (LLMs) have motivated a new generation of agent simulations in which agents can use language-like observations, reason over context, and emit structured decisions \cite{park2023generative, xi2025rise, yao2022react, hong2023metagpt}. For urban research, this direction is attractive because cities are spatially constrained, socially heterogeneous, and temporally dynamic. A useful urban agent environment should therefore expose more than a list of points of interest: it should ground agents in realistic homes, schools, workplaces, activity opportunities, infrastructure, and demographic heterogeneity.

At the same time, city-scale LLM-agent simulation faces a joint \emph{grounding--scaling} problem. On the grounding side, agents should operate under empirical urban constraints rather than in abstract text environments or coarse POI lists. On the scaling side, directly querying an LLM for every decision of every agent is impractical for hundreds of thousands of residents and many decision points per day. For example, 200,000 agents over a 24-hour day with 15-minute decision intervals would require 19.2 million online LLM calls if every decision were delegated to a teacher model. This scale mismatch motivates the central question of this paper:

\begin{quote}
\textbf{How can LLM-agent simulations be grounded in real urban constraints and scaled to city-level populations without online LLM calls for every agent decision?}
\end{quote}

We answer this question with \textbf{GenWorld}, an empirically grounded urban simulation infrastructure for scalable LLM-agent studies. The key design is a system chain rather than a single isolated algorithm: a building-level synthetic urban world provides realistic constraints; a structured agent--environment interface makes LLM-style decisions executable, validatable, and traceable; and offline compilation shifts LLM-derived decision signals out of the simulation loop into lightweight lookup policies.

\subsection{Why Urban Grounding Matters}

Urban environments are demanding testbeds for agent simulation. Physical distance and infrastructure topology shape feasible actions; schools, workplaces, and services create institutional constraints; households and neighborhoods create social and spatial context; and individual routines aggregate into city-level patterns. These characteristics make urban simulation well suited for studying situated agent behavior, but only if the environment itself is grounded in realistic population and spatial structure.

Recent LLM-agent environments often fall short on this point. Abstract text-based benchmarks support controlled reasoning tasks but lack real spatial constraints \cite{liu2023agentbench}. POI-based urban simulations and large-scale agent platforms \cite{park2023generative, piao2025agentsociety, wang2024large} have made progress in agent cognition and interaction, but they often rely on coarse spatial units, POI lists, or weakly validated population foundations. For city-scale urban studies, such abstractions make it difficult to represent building-level exposure, home--work--school anchoring, commute-distance structure, and land-use feasibility.

GenWorld therefore treats synthetic population and spatial assignment as a \emph{grounding layer}. We use established population synthesis and spatial allocation techniques, including census-constrained synthesis, building-level household placement, and school/workplace assignment. The contribution is not IPF in isolation. Rather, the contribution is integrating these standard ingredients into an LLM-ready urban world that can be used by structured agent policies and reproduced through documented data-preparation stages.

\subsection{Why Offline Compilation Matters}

LLM-agent behavior also raises a scalability problem. Online LLM inference can be useful for small numbers of agents or diagnostic traces, but city-scale rollout requires a different execution model. GenWorld separates teacher inference from simulation-time execution. A teacher LLM is queried offline under discretized context keys and finite candidate sets; the resulting score distributions are compiled into lookup policies; and simulation-time agents sample from these compiled policies while using the same validators and deterministic execution semantics.

This design trades open-ended online reasoning for tractable, auditable rollout. It should be understood as compiled stochastic policy execution rather than as fully adaptive online LLM reasoning. The benefit is that every simulated action remains tied to a structured interface, a finite candidate set, and a traceable context key, enabling large-scale execution and post-hoc analysis.

\subsection{Contributions}

This paper presents GenWorld as a reproducible infrastructure for grounded and scalable LLM-agent simulation. The contributions are organized as a connected system:

\begin{enumerate}
    \item \textbf{Empirically grounded urban world for LLM-agent simulation.} We instantiate a building-level urban world for Higashihiroshima, Japan, with 196,608 synthetic residents. Census tabulations provide demographic constraints and validation targets, while geospatial sources provide buildings, land use, POIs, roads, elevation, schools, and workplaces. We validate census consistency and use anonymized YJMob100K mobile-phone data \cite{yabe2024yjmob100k} as a commuting-distance diagnostic.

    \item \textbf{Structured agent interface and trace contract.} We define a query-conditioned interface in which raw city and persona states are mapped into binned observations, actions are selected from finite candidate sets, outputs are JSON-validated, and execution traces are recorded in machine-readable form. This contract makes LLM-style decisions executable in a simulator and suitable for later analysis or policy compilation.

    \item \textbf{Offline policy compilation for city-scale rollout.} We compile repeated teacher-model responses under discretized context keys into simulation-time lookup policies. This shifts expensive LLM calls out of the rollout loop and enables city-scale execution through lightweight sampling over validated action candidates.

    \item \textbf{Reproducible evaluation cases.} We evaluate GenWorld through three cases: a full-city weekday baseline, a weekday--weekend behavioral contrast, and a warning-response perturbation. These cases demonstrate city-scale rollout, controlled temporal-regime changes, and auditable replanning under an exogenous event. They are infrastructure diagnostics rather than calibrated forecasts of traffic, evacuation, or policy outcomes.
\end{enumerate}

\subsection{Paper Organization}

Section~\ref{sec:related_work} reviews related work in LLM-agent simulation, urban simulation platforms, synthetic population generation, and distillation for agent simulation. Section~\ref{sec:agent_interface} presents the structured agent interface and trace contract. Section~\ref{sec:distillation_scaling} describes offline policy compilation for scalable rollout. Section~\ref{sec:grounding} details the empirically grounded urban world construction and validation. Section~\ref{sec:architecture} presents the platform architecture and simulation engine. Section~\ref{sec:results} reports the evaluation cases and scalability analysis. Sections~\ref{sec:discussion} and~\ref{sec:conclusion} discuss limitations and conclude.

\section{Related Work}
\label{sec:related_work}

Table \ref{tab:comparison} provides an overview of how GenWorld compares to existing platforms across three categories: LLM agent simulation platforms, LLM-based urban mobility platforms, and population synthesis platforms. We detail these comparisons in the following subsections.

\begin{table*}[t]
\centering
\small
\setlength{\tabcolsep}{4pt}
\renewcommand{\arraystretch}{1.05}
\caption{Comparison of GenWorld with Related Platforms}
\resizebox{\textwidth}{!}{%
\begin{tabular}{@{}lcccccc@{}} 
\toprule
\textbf{Platform} & \textbf{Population} & \textbf{Empirical} & \textbf{Scale} & \textbf{Real} & \textbf{Spatial} & \textbf{Social} \\
 & \textbf{Realism} & \textbf{Validation} & \textbf{(Agents)} & \textbf{Geography} & \textbf{Detail} & \textbf{Networks} \\
\midrule
\multicolumn{7}{l}{\textit{LLM Agent Simulation Platforms}} \\
GridWorld/TextWorld & Low & No & $<100$ & No & No & No \\
Generative Agents \cite{park2023generative} & Low & No & $<100$ & No & Limited & Limited \\
WebArena \cite{zhou2023webarena} & N/A & N/A & Individual & No & No & No \\
\midrule
\multicolumn{7}{l}{\textit{LLM-Based Urban Mobility Platforms}} \\
LLM-ABM Framework \cite{liu2025toward} & Low & No & $<100$ & No & Low & No \\
LLMob \cite{wang2024large} & Medium & GPS & Individual & Yes & POI-level & No \\
TrajLLM \cite{ju2025trajllm} & Medium & Qualitative & $<100$ & No & POI-level & No \\
MobAgent \cite{li2024more} & Medium & Survey & Individual & Yes & POI-level & No \\
GATSim \cite{liu2025gatsim} & Medium & No & 1K--10K & No & Medium & Limited \\
MobileCity \cite{ye2025mobilecity} & Medium & No & 1K--10K & No & Medium & Limited \\
OpenCity \cite{yan2024opencity} & Low & GPS & 1K--10K & Yes & POI-level & No \\
\midrule
\multicolumn{7}{l}{\textit{Population Synthesis Platforms}} \\
Jiang et al. \cite{jiang2022method} & High & Census & 100K+ & Yes & Road-based & Multi-layer \\
Pseudo-PFLOW \cite{kashiyama2024nationwide} & High & Census & 100K+ & Yes & Building & No \\
\midrule
\textbf{GenWorld (Ours)} & \textbf{High} & \textbf{Multi-source} & \textbf{100K+} & \textbf{Yes} & \textbf{Building} & Multi-layer$^\dagger$ \\
\bottomrule
\multicolumn{7}{l}{\footnotesize $^\dagger$ Social networks are generated from spatial co-location but not used in current experiments.}
\end{tabular}
}
\label{tab:comparison}
\end{table*}

\subsection{LLM Agents and Simulation Platforms}

 The emergence of Large Language Models has driven rapid progress in autonomous agent systems. Recent works demonstrate LLM agents across a range of settings, from social simulation \cite{park2023generative} to tool use \cite{schick2023toolformer} and multi-agent collaboration \cite{hong2023metagpt}. This progress motivates the need for realistic simulation environments that can support LLM agent research under real-world constraints.

\noindent\textbf{Existing Agent Platforms.} Existing platforms and benchmarks span multiple levels of realism.

\textbf{Abstract environments} (e.g., GridWorld/TextWorld-style tasks) \cite{liu2023agentbench} are useful for isolating reasoning and planning, but they abstract away geography, resource constraints, and social interactions.

\textbf{Task-specific platforms} such as SWE-bench \cite{jimenez2023swe} (software engineering) and WebArena \cite{zhou2023webarena} (web navigation) provide grounded objectives and measurable success criteria, but they typically focus on single-agent, non-spatial settings.

 \textbf{Social simulation platforms} such as Generative Agents \cite{park2023generative} explore emergent interactions, yet the environments are simplified and the scale (e.g., 25 agents) is insufficient for studying city-scale phenomena and computational scalability. CityBench~\cite{feng2024citybench} evaluates LLM world-modeling capabilities for urban tasks but does not provide building-level population grounding.

 \noindent\textbf{LLM Agents in Transportation and Mobility.} Beyond interactive simulacra, LLMs have been explored as simulated economic agents~\cite{horton2023large} and integrated into mobility and transportation settings. LLMob \cite{wang2024large} uses self-consistency and retrieval-augmented strategies for individual mobility generation with GPS-based validation. Liu et al.~\cite{liu2025toward} outline an LLM-agent-based transportation modeling framework with a small proof-of-concept. TrajLLM~\cite{ju2025trajllm} combines LLM-based persona generation with hybrid destination choice (LLM + physical models), but focuses on POI-level trajectories. GATSim~\cite{liu2025gatsim} and MobileCity~\cite{ye2025mobilecity} target larger-scale mobility simulation; MobileCity achieves efficiency partly by disabling LLM modules at scale, trading behavioral fidelity for speed. OpenCity~\cite{yan2024opencity} proposes a ``group-and-distill'' prompt optimization strategy that clusters agents with similar attributes and distills shared reasoning patterns, achieving 600$\times$ acceleration in simulation time; however, it focuses on prompt-level efficiency rather than building-level spatial grounding. Overall, these efforts primarily emphasize individual trajectory generation or engineering efficiency. They often do not provide city-scale population synthesis with jointly validated demographics and spatial assignments (e.g., building-level placement) or thorough empirical validation.

 Existing platforms often do not jointly provide realistic population foundations supported by empirical data, spatial complexity with infrastructure constraints, computational scalability to city-scale (100,000+ agents), and LLM-compatible interfaces. GenWorld provides an empirically grounded urban environment with 200,000-agent scalability based on data from Higashihiroshima, Hiroshima, Japan.

\subsection{Urban Simulation Platforms}
\label{sec:related_platforms}

Agent-based modeling has a rich history in urban and transportation research~\cite{epstein1996growing, lazer2009computational}, with several established platforms:

\noindent\textbf{Traditional ABM Platforms.} \textbf{GAMA} \cite{taillandier2019building}, \textbf{MASON} \cite{luke2005mason}, and \textbf{NetLogo} \cite{tisue2004netlogo} are widely used for urban simulation. These platforms provide powerful modeling capabilities but were designed for domain experts rather than AI researchers, and they do not provide standardized LLM integration interfaces or natural language observation spaces.

\noindent\textbf{Transportation Simulation Tools.} MATSim \cite{horni2016introducing}, SUMO \cite{krajzewicz2012recent}, and similar tools focus on traffic simulation with detailed traffic modeling. However, they typically use simplified behavioral models and do not incorporate the cognitive realism enabled by LLM-driven agents.

\noindent\textbf{Commercial Platforms.} AnyLogic, Citilabs, and other commercial tools offer sophisticated urban modeling but are closed-source, expensive, and not designed for AI research integration.

 Recent open-source efforts such as VoxCity~\cite{fujiwara2025voxcity} provide seamless 3D urban environment generation, while Biljecki and Chow~\cite{biljecki2022global} establish global building morphology indicators for standardized urban analysis. However, existing platforms were not designed with LLM agents in mind. GenWorld aims to address these gaps by providing natural language observation spaces, flexible action specifications, validated population foundations, and computational scalability through knowledge distillation.

\subsection{Synthetic Population Generation}

Generating realistic synthetic populations is fundamental to valid agent-based modeling \cite{lim2025large}.

 \noindent\textbf{Population Synthesis Methods.} \textbf{Iterative Proportional Fitting (IPF)} \cite{choupani2016population} and its variants are commonly used methods, adjusting cell weights to match marginal distributions from census data. Beyond IPF, prior work also explores alternative formulations such as combinatorial optimization, Bayesian approaches, and deep generative models (DGMs). While DGMs can generate diverse populations beyond observed samples, they often struggle to balance \textit{sampling zeros} (valid but unobserved combinations) with \textit{structural zeros} (implausible combinations) \cite{lim2025large}. Recent work explores LLM-based approaches: Li et al.~\cite{li2024more} proposed MobAgent, using LLMs to extract fine-grained mobility patterns from individual profiles through self-evaluation and recursive reasoning, validated on 0.2M travel surveys. Ma et al.~\cite{ma2025learning} developed a foundation model using LLMs for semantic enrichment of GPS trajectories, demonstrating transfer learning across regions (LA to Egypt) for mobility pattern synthesis. While these LLM-based methods have been explored for individual trajectory generation, they focus on personal mobility modeling rather than city-scale population synthesis with validated demographic distributions and spatial assignments.

  \noindent\textbf{Spatial Assignment and Social Networks.} Assigning synthetic individuals to geographic locations is important for spatial realism. Common approaches include: \textbf{gravity models} \cite{anderson2011gravity} for workplace assignment, \textbf{distance-based allocation} for household placement, and \textbf{constraint satisfaction} for student-to-school assignment. Jiang et al.~\cite{jiang2022method} developed a large-scale method generating 23 million geographically-explicit individuals for New York Metro Area with multi-layer social networks (household, work, school, daycare) emergent from spatial co-location, highlighting the importance of social networks for urban simulations. Kashiyama et al.~\cite{kashiyama2024nationwide} developed Pseudo-PFLOW, an agent-based framework that downscales census data to building-level assignments using Markov chain models for activity generation, covering Japan's 130 million population. While achieving strong validation results (R$^2$=0.61--0.98 for population distribution), these approaches rely on traditional statistical models rather than LLM-driven behavioral realism and lack integration with modern LLM agent frameworks.

  \noindent\textbf{Validation Approaches.} Traditional validation relies primarily on census data comparison. Recent work has begun incorporating \textbf{mobile phone data} \cite{yabe2024yjmob100k} for validating commuting patterns, building on foundational studies of human mobility patterns~\cite{gonzalez2008understanding, pappalardo2018data, hasan2013spatiotemporal}. Ma et al.~\cite{ma2025learning} demonstrated multi-level validation through traffic simulation, achieving MAPE $<$ 6\% for traffic volumes. However, systematic validation combining demographic distributions, spatial assignments, and mobility patterns against real-world data remains rare.

 Most synthetic population studies focus on demographic accuracy but neglect spatial validation with real mobility data, social network construction, daily activity schedules, and integration with LLM agent frameworks. GenWorld provides an end-to-end pipeline that covers these aspects.

\subsection{Knowledge Distillation for Agent Simulation}

Knowledge distillation \cite{hinton2015distilling} has been widely applied in machine learning to compress large models into efficient ones. Recent applications include:

Beyond model compression, recent work explores abstraction and software architecture to scale LLM-agent simulations. Chopra et al.~\cite{chopra2024limits} introduce \emph{LLM archetypes}, where many agents share an archetypal LLM policy to increase throughput at scale, but this can reduce individual-level heterogeneity and online adaptivity. SocioVerse~\cite{zhang2025socioverse} targets population-scale social simulation by aligning LLM agents to a large pool of real users and standardizing simulation procedures; however, it relies on large external datasets and its alignment pipeline can be costly to reproduce or transfer. For influence diffusion in social networks, LLM-AIDSim~\cite{zhang2025llm} integrates LLM-enhanced agents into classical diffusion simulation pipelines, but the approach is task-specific and may not directly generalize to open-ended urban decision spaces. From a systems perspective, SALLMA~\cite{becattini2025sallma} proposes a layered multi-agent architecture with orchestration and containerized deployment; while improving modularity and scalability, it does not inherently remove per-decision LLM inference costs and can require substantial engineering infrastructure.

\noindent\textbf{LLM Distillation.} Distilling large language models into smaller, faster models while maintaining performance is an active area of research. However, most work focuses on natural language tasks, not agent decision-making in complex environments.

\noindent\textbf{Agent Behavior Cloning.} Imitation learning and behavior cloning train efficient policies from expert demonstrations. GenWorld extends this paradigm by using LLMs as "expert demonstrators" to generate training data for efficient student models.

 We apply knowledge distillation to enable city-scale LLM agent simulation. Our approach estimates the teacher's discrete decision distribution via repeated Monte Carlo sampling and compiles the resulting probabilistic policy into efficient lookup tables, shifting expensive inference out of the simulation loop and enabling large speedups in typical settings for large-scale simulations.

 As summarized in Table \ref{tab:comparison}, GenWorld combines \textbf{building-level population grounding} with census-validated demographics, \textbf{city-scale scalability} via offline knowledge distillation (200,000+ agents), \textbf{multi-layer social networks} derived from spatial co-location, and \textbf{schema-validated LLM-ready interfaces} that produce machine-readable behavioral traces in a real-city instantiation.

\section{Agent Interface}
\label{sec:agent_interface}

GenWorld exposes a lightweight decision interface for LLM agents and records each decision as a structured log entry. This interface is designed to enable post-hoc qualitative inspection of agent routines and failure modes and provide machine-readable decision traces for offline distillation. Concretely, each decision consumes a binned observation $\tilde{o}_{i,t}$ and a finite candidate set $\mathcal{A}_{i,t}$, and produces a schema-conformant JSON action, a validator bit, and (if needed) a deterministic fallback outcome, all recorded as a log entry.

\begin{figure*}[t]
     \centering
     \includegraphics[width=0.98\textwidth]{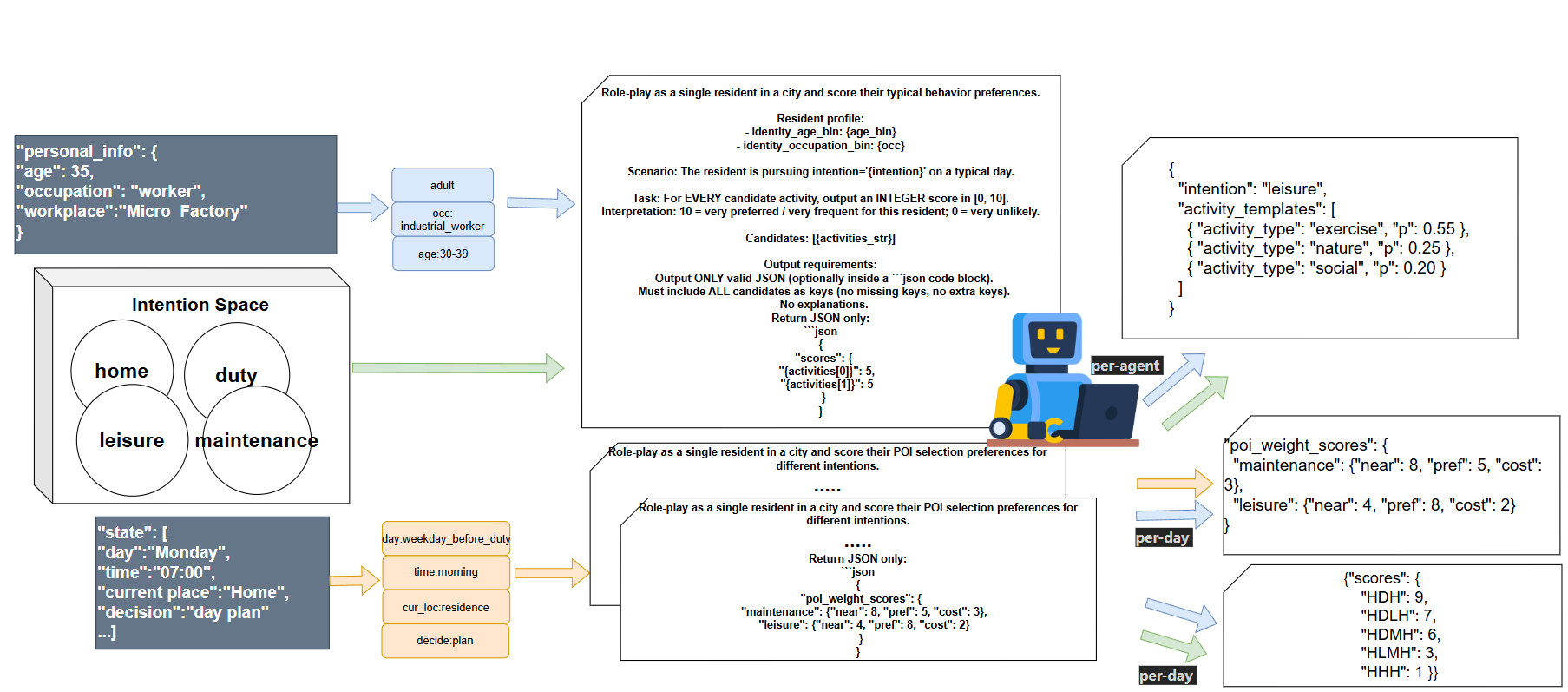}
     \caption{Query-conditioned prompt construction for our structured decision interface. Raw persona/state fields are deterministically mapped into coarse bins and are not included verbatim in the prompt. The figure schematically illustrates prompt variants used in this instantiation: a per-agent \texttt{ActivityPreference} query over a fixed candidate set under a given intention, and day-level prompts that score POI-selection preferences over \texttt{near/pref/cost} weights and intention-chain templates over a predefined chain candidate set. In this default instantiation, POI-weight scoring and intention-chain scoring are issued jointly as a single \texttt{DayPlan} query, but they can also be queried separately. Input features are represented using coarse discrete bins, while candidate scores returned by the teacher are integers in $[0,10]$ over a predefined option set. Section~\ref{sec:distillation_scaling} describes how these structured traces are aggregated and compiled for scalable rollout.}
     \label{fig:bin2prompt}
\end{figure*}

\begin{figure}[htbp]
     \centering
     \includegraphics[width=\columnwidth]{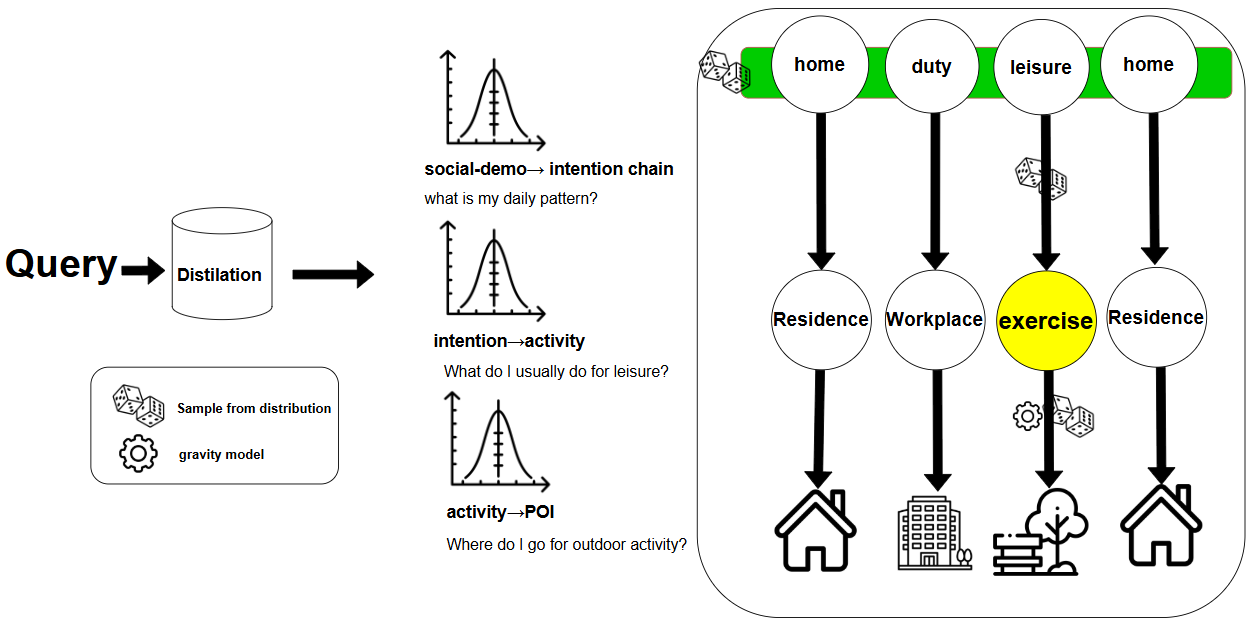}
     \caption{Plan-to-trajectory execution with a two-tier decision structure. \texttt{ActivityPreference} provides persona-conditioned activity propensities, while \texttt{DayPlan} specifies intention-chain templates and POI-selection weights. A lightweight executor produces explicit trajectories through fixed ontologies and feasibility checks.}
     \label{fig:prompt2traj}
\end{figure}

\paragraph{Observation and Action Schema}
At each decision point for agent $i$ at time $t$, the simulator constructs a decision context from the city state $x_t$ (time, environment signals, and infrastructure states), a synthesized persona $u_i$ produced by the population instantiation pipeline (core demographics and spatial anchors such as home/work/school when available, with optional household and social features), and optionally short-term memory summaries $m_{i,t}$ distilled from recent logs. This context is denoted as $c_{i,t}=(x_t, u_i, m_{i,t})$. Given a decision query $q_t$, the environment deterministically produces a binned observation and a finite candidate action set:
\begin{align*}
 \tilde{o}_{i,t} &= \phi(c_{i,t}; q_t),\\
 \mathcal{A}_{i,t} &= \kappa(q_t, \tilde{o}_{i,t}).
\end{align*}
The function $\phi$ is implemented as a deterministic encoder stack that includes coarse binning and query-specific formatting. A prompt composer $g(\tilde{o}_{i,t}, q_t)$ assembles a stable template with question-specific slots. The agent then outputs a structured JSON action $a_{i,t}\in\mathcal{A}_{i,t}$ following a fixed schema (e.g., activity type). A deterministic validator $v(\tilde{o}_{i,t}, a_{i,t})\in\{0,1\}$ enforces schema and feasibility constraints; invalid actions trigger a deterministic safe fallback before execution, and all artifacts are logged.
Figure~\ref{fig:bin2prompt} illustrates a representative query where raw persona/state fields are deterministically mapped into coarse bins before being passed to the LLM. Figure~\ref{fig:prompt2traj} summarizes how the resulting structured outputs are executed into full-day trajectories by lightweight deterministic rules.
In this implementation, persona slices are intentionally sparse, while richer preference/trait slices can be added as optional extensions or treated as latent variables depending on the target application.

\paragraph{Two-Tier Decision Queries for Long-Horizon Rollout}
Decision-making is separated into two structured outputs with different time scales. \texttt{ActivityPreference} is a per-agent, persona-conditioned preference profile that is initialized once (and optionally refreshed) and defines propensities over activity types for each high-level intention. \texttt{DayPlan} is a per-day (or per-checkpoint) plan that specifies a small mixture of intention-chain templates together with discretized POI-selection weights. The plan sampling index is denoted by $k$ (day-start or checkpoint), which is much sparser than the simulator time step $t$ used for execution, and the city state at plan sampling time is written as $x_k$. The intention space is fixed to \{\texttt{home}, \texttt{duty}, \texttt{leisure}, \texttt{maintenance}\}.

This two-tier abstraction is grounded in time-geography theory \cite{hagerstrand1970people}: daily mobility is constrained by capability (physical limits), coupling (coordination with others), and authority (institutional schedules). Our intention hierarchy (home/duty/leisure/maintenance) captures these canonical constraint classes, while the activity vocabulary covers the primary purposes observed in national time-use surveys. The fixed ontology trades open-ended expressiveness for tractability and repeatability; extending the vocabulary is straightforward within the same interface contract.

Critically, the day-level query is not conditioned on a single intention; instead, the simulator provides a small, day-type-specific candidate set of intention-chain templates (e.g., weekday vs.\ weekend variants) and includes this candidate set as part of the binned context.
During rollout, agents sample a \texttt{DayPlan} at day start, and the simulator consumes it through a lightweight executor (as shown in Figure~\ref{fig:prompt2traj}) to produce an explicit trajectory of simulator actions. Concretely, an intention-chain template is sampled from the day-type-specific candidate set, expanded into activity types by sampling \texttt{ActivityPreference}, and grounded into concrete destinations via a fixed activity-to-place ontology and feasibility checks. Overrides may be requested by the agent or forced by the simulator when feasibility checks fail or exogenous events invalidate the plan. In both cases, a deterministic return-home fallback is applied and the agent stays at home until the next plan sampling time (day-start or checkpoint). Section~\ref{sec:distillation_scaling} describes how decision traces are collected under binned contexts and compiled into scalable student policies.

\paragraph{Formal Contract Summary}
The formal contract (Figure~\ref{fig:bin2prompt}) is summarized as follows. The simulator deterministically maps raw persona and state fields into coarse bins via encoders $b_u$ and $b_x$:
\begin{align*}
\mathcal{I} &= \{\texttt{home},\texttt{duty},\texttt{leisure},\texttt{maintenance}\},\\
\tilde{u}_i &= b_u(u_i),\quad \tilde{x}_k = b_x(x_k),\\
\tau &\in\{\texttt{weekday},\texttt{weekend}\},\\
\mathcal{C}_{\tau} &\subseteq \mathcal{I}^{*}.
\end{align*}
where $\tau$ is a coarse day-type label and $\mathcal{C}_{\tau}$ is a small predefined candidate set of intention-chain templates. The per-agent query defines a conditional categorical distribution over activity types given intention $z$:
\begin{align*}
A_i(z) &:= \texttt{ActivityPreference}_i(z),\\
A_i(z) &= \{(a, p(a\mid z))\}_{a\in\mathcal{A}_z},\quad z\in\mathcal{I},\\
a &\in\mathcal{A}_z,\quad \sum_{a\in\mathcal{A}_z} p(a\mid z) = 1,
\end{align*}
where $\mathcal{A}_z$ is a small predefined set of activity types allowed under intention $z$ (Appendix~\ref{app:intention_activity_taxonomy}). The per-day (or per-checkpoint) query returns:
\begin{align*}
D_{i,k} &= \texttt{DayPlan}_{i,k}(\tilde{x}_k, \tilde{u}_i, \mathcal{C}_{\tau}),\\
D_{i,k} &= (r_{i,k}, C_{i,k}, w_{i,k}),\\
C_{i,k} &= \{(c_j, \pi_j)\}_{j=1}^{|\mathcal{C}_{\tau}|},\\
r_{i,k} &\in\{0,1\},\quad \sum_{j=1}^{|\mathcal{C}_{\tau}|} \pi_j = 1,\\
c_j &\in \mathcal{C}_{\tau}\subseteq \mathcal{I}^{*},\\
w_{i,k}(z) &= (\ell^{\texttt{near}}_{i,k}(z),\ell^{\texttt{pref}}_{i,k}(z),\ell^{\texttt{cost}}_{i,k}(z)),\\
\ell^{*}_{i,k}(z) &\in \{0,\dots,10\},\quad z\in\mathcal{I},
\end{align*}
where $\mathcal{C}_{\tau}$ is a small predefined candidate set of intention-chain templates (in our instantiation, $|\mathcal{C}_{\tau}|=6$ per day type). Here $r_{i,k}$ is an override request flag, each $c_j$ is an intention-chain template, and $w_{i,k}(z)$ specifies discretized POI-selection weights for intention $z$. Here $k$ denotes the plan sampling index (day-start or checkpoint), which is much sparser than the execution time step. Overrides may also be forced by the simulator when feasibility checks fail; in either case, a deterministic return-home fallback is applied.

\paragraph{Tool-Oriented Interface, Robustness, and Traceability}
The interface is realized as stable prompt templates with strict JSON schemas that are validated and logged by the simulator, and can be wrapped by standard tool-calling middleware when needed.
A fixed, query-conditioned observation schema, a discrete and bounded action space with strict validation, and deterministic execution semantics are enforced.
At city scale, even rare formatting or parsing failures can derail long simulations. LLM decisions are therefore constrained to a small discrete action set with a fixed schema, and strict validation and deterministic fallback rules are enforced in the decision logger. This design makes decision traces directly machine-readable and suitable for downstream analysis and policy compilation (Section~\ref{sec:distillation_scaling}).

\section{Distillation and Scaling}
\label{sec:distillation_scaling}

To scale LLM-driven decision-making to city-scale simulations, the teacher's stochastic decision behavior is distilled into empirical score vectors and sampling distributions under discretized contexts by repeatedly querying the LLM under identical context keys and aggregating its scores over a fixed candidate set (e.g., intention-chain templates or intention-conditioned activity templates). Because the interface bins raw contexts into discrete keys and restricts each query to a finite candidate set with strict validation, the teacher can be repeatedly queried under identical keys and its scores can be aggregated. The key idea is to shift expensive inference out of the simulation loop: a one-time offline cost is paid to estimate these distributions, and the resulting compiled tables are executed via amortized constant-time lookup and sampling given bounded candidate sets per query, with respect to the number of agents and decision steps.

In a micro-benchmark on the compiled \texttt{ActivityPreference} table, Python lookup achieves 1.85M queries/s (0.54\,$\mu$s per query) over 200,000 randomized context keys. While absolute throughput depends on hardware and implementation details, this benchmark highlights the potential for large speedups relative to online LLM inference in typical settings. End-to-end wall-clock time per simulator step also includes environment updates, routing, and execution overheads. Prompt templates used for distillation are listed in Appendix~\ref{app:distillation_prompts}.

\begin{figure*}[t]
     \centering
     \includegraphics[width=0.98\textwidth]{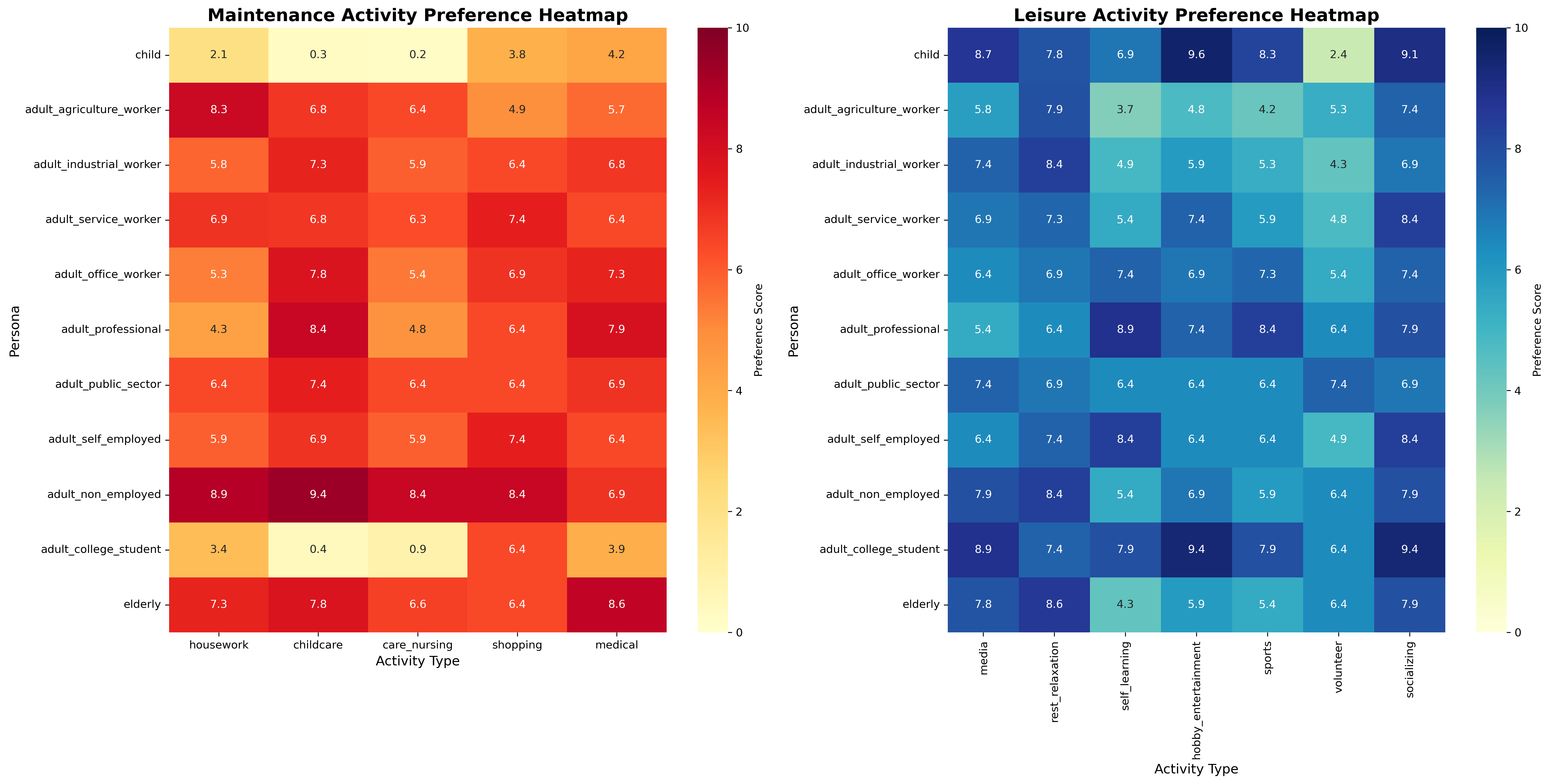}
     \caption{Teacher preference scores (0--10) for \texttt{ActivityPreference} across persona categories (rows) and candidate activity types (columns), shown separately for maintenance (left) and leisure (right). The scores define the simulation-time sampling distribution used by the compiled policy.}
     \label{fig:activity_preference_heatmap}
\end{figure*}

\begin{figure*}[t]
     \centering
     \begin{subfigure}[b]{0.46\textwidth}
           \centering
           \includegraphics[width=\textwidth]{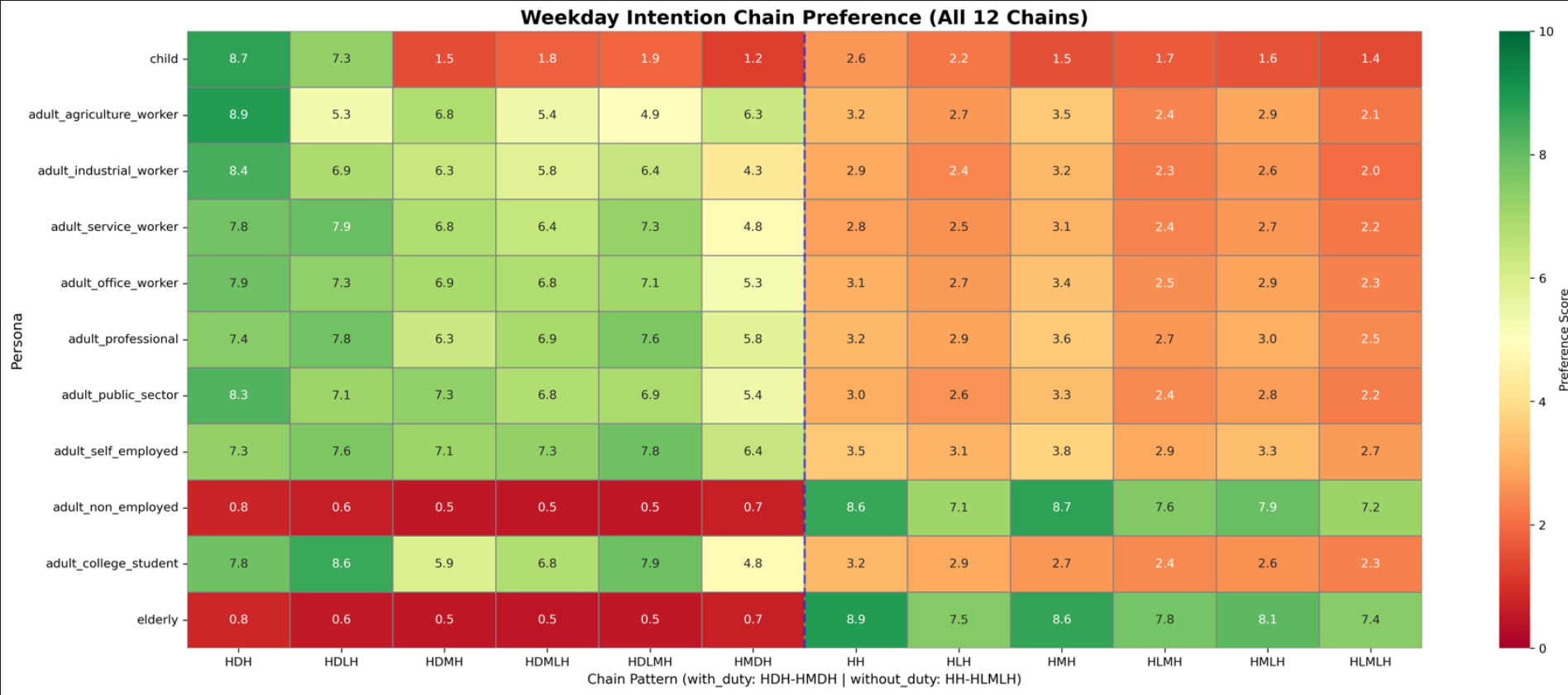}
           \caption{Weekday intention-chain template preference.}
     \end{subfigure}
     \hfill
     \begin{subfigure}[b]{0.46\textwidth}
           \centering
           \includegraphics[width=\textwidth]{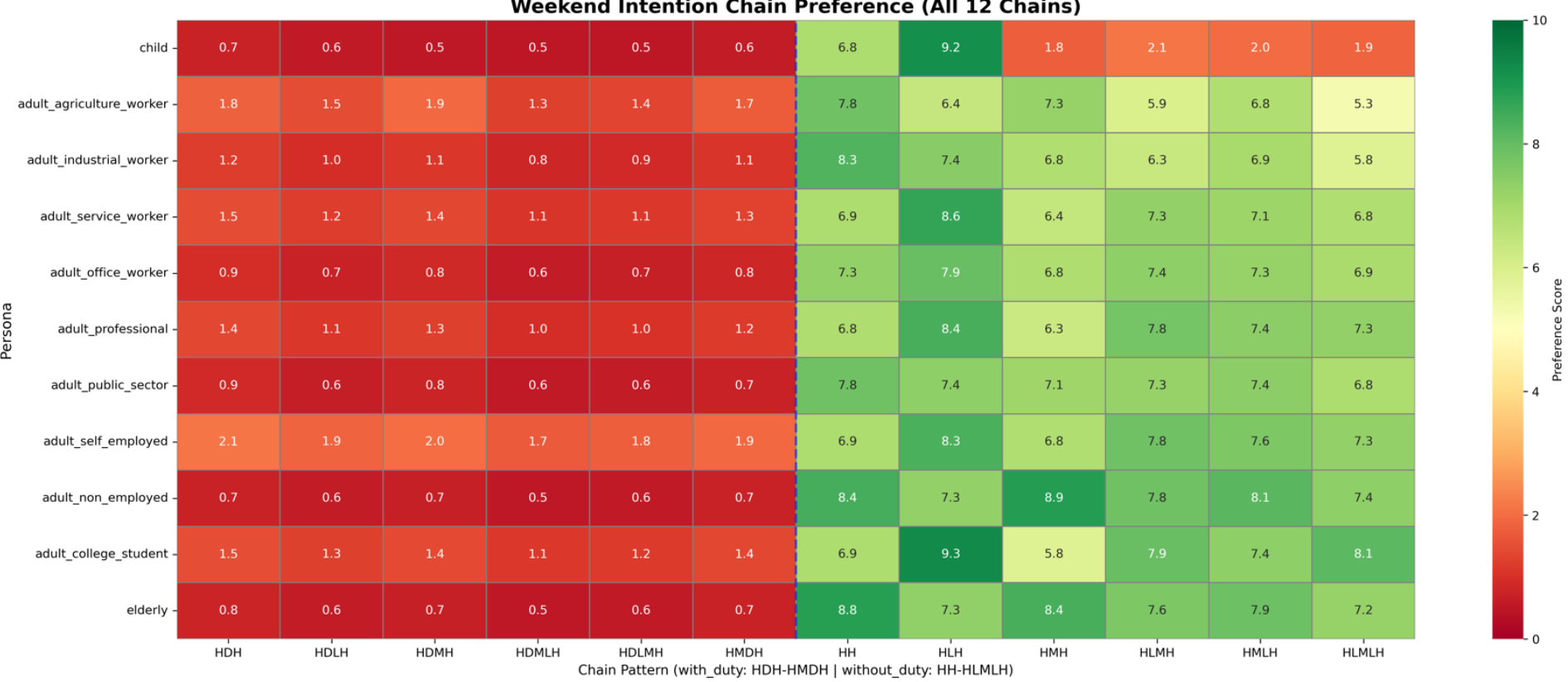}
           \caption{Weekend intention-chain template preference.}
     \end{subfigure}
     \caption{Distilled teacher scores for \texttt{DayPlan} intention-chain templates, shown separately for weekday and weekend candidate sets.}
     \label{fig:dayplan_pattern}
\end{figure*}

\paragraph{Action Primitives and Context Discretization}
Repeated sampling requires that the teacher be queried under identical contexts. Following the interface contract in Section~\ref{sec:agent_interface}, raw persona and state are discretized into bins (e.g., $\tilde{u}_i=b_u(u_i)$, $\tilde{x}=b_x(x)$) and each decision query $q_t$ is treated as defining its own finite action space. Concretely, for each query type $q_t$ (e.g., \texttt{ActivityPreference} or \texttt{DayPlan}), an executable discrete action set $\mathcal{A}_{q_t}$ is defined that matches the simulator's structured schema and validation rules.
A finite context key $s=(\tilde{u}_i,\tilde{x},q_t,\tau)$ is then formed, where $\tau$ indexes the day-type-specific candidate template set used by \texttt{DayPlan}. This makes repeated offline teacher aggregation well-defined and enables compilation into amortized constant-time lookup policies.
The day-type indicator $\tau$ is included explicitly because the \texttt{DayPlan} candidate set differs across day types (e.g., weekday vs.\ weekend).

\paragraph{Computational Motivation}
At city scale, a direct teacher-driven simulation requires $O(NT)$ LLM calls, where $N$ is the number of agents and $T$ is the number of decision points per simulated day. For example, $N=200{,}000$ agents with 15-minute time steps over 24 hours yields $T=96$ and thus $1.92\times 10^7$ calls for a single day, which is computationally expensive in practice. Distillation reduces simulation-time inference to amortized constant-time table lookup and sampling with respect to the number of agents and decision steps.

\paragraph{Repeated Teacher Query Aggregation}
For a fixed candidate set $\mathcal{A}_{q_t}$ and context key $s$, $K$ teacher score vectors $\{r^{(k)}(\cdot)\}_{k=1}^{K}$ are sampled, where each query returns an integer score $r^{(k)}(a)\in[0,10]$ for every candidate $a\in\mathcal{A}_{q_t}$. The mean score and consistency statistics are aggregated:
\begin{align}
\mu(a\mid s) &= \frac{1}{K} \sum_{k=1}^{K} r^{(k)}(a),\\
\sigma(a\mid s) &= \sqrt{K^{-1}\sum_{k=1}^{K} \bigl(r^{(k)}(a)-\mu(a\mid s)\bigr)^2}.
\end{align}
The aggregated mean scores are normalized across candidates into a categorical sampling distribution $\pi(\cdot\mid s)$, which is used for simulation-time sampling.
Since scores are in $[0,10]$, an executable sampling distribution is constructed by normalizing the mean scores:
\begin{align}
\pi(a\mid s) &= \mathrm{Normalize}\left(\mu(a\mid s)\right),\quad a\in\mathcal{A}_{q_t}.
\end{align}
The score variability $\sigma(a\mid s)$ is reported to quantify teacher consistency across repeated queries (and to diagnose context regions with high variability).

\paragraph{Policy Compilation and Simulation-Time Inference}
The aggregated scores and sampling distributions (e.g., $\mu(\cdot\mid s)$ and $\pi(\cdot\mid s)$) are compiled into per-query lookup tables keyed by discretized context features (persona bins, time bins, coarse location types, scenario indicators, and day-type indicators). During simulation, agents sample an intention-chain template or activity template according to the distilled distribution rather than querying the LLM, and execute the sampled schema through the same validator/executor as the teacher outputs. Scoring and sampling over intention-chain templates enables long-horizon diversity while keeping the execution interface lightweight. This compilation separates two concerns:
\begin{itemize}
    \item \textbf{Teacher inference (offline)}: generate multiple samples per context to estimate $\mu(\cdot\mid s)$ (and $\sigma(\cdot\mid s)$), then derive $\pi(\cdot\mid s)$.
    \item \textbf{Agent rollout (online)}: execute a lightweight stochastic decision rule by table lookup and sampling.
\end{itemize}

\paragraph{Context Design and Coverage}
To make compilation feasible, contexts are discretized into a finite key space (e.g., persona bins, coarse location types, and time bins) and representative contexts are sampled according to the instantiated population distribution. This allows the offline sampling budget to be allocated where it matters most while keeping simulation-time inference amortized constant-time with respect to the number of agents and decision steps. This discretization trades off fidelity for tractability: behavior matching depends on context key design and coverage, and unseen keys may require backing off to coarser keys or a conservative default.

\section{Empirical Grounding of the Urban World}
\label{sec:grounding}

Our reference instantiation integrates multi-source empirical datasets, including official census statistics and administrative boundaries, building footprints and POIs, parcel-level land-use labels, a complete road network with elevation, and anonymized mobility data for commuting diagnostics. These inputs provide constraints for population synthesis and spatial grounding, and also provide independent signals for validation.

Detailed data sources and processing steps are provided in Appendix~\ref{app:data_sources} (Table~\ref{tab:data_sources}).

\subsection{Population and Environment Foundation}
\label{subsec:population_env}

This section describes the empirically-grounded population and environment foundation used in our Higashihiroshima reference instantiation, grounded in publicly available census tabulations and geospatial layers (buildings, land use parcels, school districts, POIs, and roads), synthesizing 196,608 individuals across 89,988 private households, with 5,641 additional group-quarter records retained separately. The formulation combines demographic micro-synthesis under tract-level census constraints with spatial grounding of home, school, and work locations under capacity and distance constraints.

\subsubsection{Tract-Level Micro-Synthesis and Attribute Assignment}
For each tract $t$, the total population $N_t$ is treated as a hard constraint and an age--gender joint distribution is estimated whose marginals match census age counts and gender totals. A 2D IPF procedure is adopted on an age\,$\times$\,gender matrix $M^{(t)}$:
\begin{align}
M_{a,g}^{(k+\frac{1}{2})} &= M_{a,g}^{(k)} \cdot \frac{n_{t,a}}{\sum_{g'} M_{a,g'}^{(k)}} \\
M_{a,g}^{(k+1)} &= M_{a,g}^{(k+\frac{1}{2})} \cdot \frac{n_{t,g}}{\sum_{a'} M_{a',g}^{(k+\frac{1}{2})}} \label{eq:ipf}
\end{align}
where $n_{t,a}$ is the census count of age bin $a$ in tract $t$, and $n_{t,g}$ is the census total of gender $g\in\{\text{male},\text{female}\}$. $M^{(t)}$ is initialized with a strictly positive prior (e.g., uniform or tract-independent) and Eq.~\eqref{eq:ipf} is iterated until marginal errors fall below $\epsilon$ or for a fixed number of rounds. Individuals are then sampled from the normalized joint distribution, and a concrete integer age is sampled uniformly within the selected age bin.

Given the sampled individuals, households are formed using the tract household-size histogram (1,2,3,4,5,6+) with a lightweight plausibility heuristic (e.g., capping household size at 6). The census household count target $H_t$ is enforced and household sizes are sampled to match the tract histogram. Employment status and occupation categories are then assigned for working-age individuals so that tract-level employed totals and occupational marginals match the census. Let $\mathcal{I}_t$ be individuals in tract $t$, and $\mathcal{W}_t\subset\mathcal{I}_t$ be working-age individuals. Denote the census employed target as $E_t$ and the census occupation target counts as $C_{t,o}$ for occupation category $o\in\mathcal{O}$. Let $E_t' = \min(E_t,\,|\mathcal{W}_t|)$ and let $C'_{t,o}$ be adjusted occupation targets derived from $\{C_{t,o}\}_{o\in\mathcal{O}}$ by padding/truncation so that $\sum_{o\in\mathcal{O}} C'_{t,o} = E_t'$. The following constraints are enforced:
\begin{align}
\sum_{i\in\mathcal{W}_t} \mathbb{I}[\text{employed}_i] &= E_t' \label{eq:stage3_employ}\\
\sum_{i\in\mathcal{W}_t} \mathbb{I}[\text{employed}_i \land \text{occ}_i=o] &= C'_{t,o},\quad \forall o\in\mathcal{O} \label{eq:stage3_occ}
\end{align}
Eq.~\eqref{eq:stage3_employ}--\eqref{eq:stage3_occ} are realized via seeded sampling: an employed subset of size $E_t'$ is drawn and an occupation multiset with counts $C'_{t,o}$ is assigned, followed by a tract-seeded random permutation.

\subsubsection{Spatial Grounding of Home, School, and Work}
Households are assigned to residential buildings within each tract using a capacity-aware allocation; students are assigned to schools using district polygons when available with nearest-school fallback; university assignment uses a distance-based stochastic choice with weights proportional to $1/d^2$. For workplace allocation, employed individuals are mapped to \textit{landuse} parcels (not building IDs) using an occupation-conditioned landuse prior (occupation$\rightarrow$landuse mapping with ratios $r_{o,l}$) together with a maximum commute-distance constraint $d_{\max}$.

\emph{Capacity inference by area (quotas).} For occupation $o$, let total employees be $N_o$, eligible landuse categories be $\mathcal{L}_o$, and the configured landuse ratio be $r_{o,l}$ for $l\in\mathcal{L}_o$ with $\sum_{l\in\mathcal{L}_o} r_{o,l}=1$. For each landuse parcel $j$ of category $l$ with area $A_j$, an occupation-specific quota is defined:
\begin{align}
q_{j,o} &= \frac{N_o\, r_{o,l}\, A_j}{\sum_{k\in\mathcal{P}_l} A_k},\quad j\in\mathcal{P}_l,\ l\in\mathcal{L}_o \label{eq:stage5_quota}
\end{align}
where $\mathcal{P}_l$ is the set of parcels with landuse category $l$.
Fractional quotas are converted into integer capacities $\hat{q}_{j,o}$ (e.g., via floor with remainder redistribution or stochastic rounding) to preserve total capacity per occupation.

\emph{Gravity-based allocation.} We employ a gravity model to assign workplaces, balancing employment opportunities with distance decay. Let $d_{ij}$ be the haversine distance between employed individual $i$'s home and landuse parcel $j$. The probability $P_{ij}$ of individual $i$ choosing workplace $j$ is proportional to the parcel's destination attractiveness (capacity) and inversely proportional to commute distance:
\begin{align}
P_{ij} \propto A_j^{\alpha} \cdot f(d_{ij}) \cdot M_{ij} \label{eq:gravity_prob}
\end{align}
where $A_j$ is the capacity (attractiveness) of parcel $j$, $f(d) = d^{-\beta}$ is the distance decay function with friction parameter $\beta$, and $M_{ij}$ is a binary mask enforcing occupation compatibility ($M_{ij}=1$ if parcel $j$ supports individual $i$'s occupation $o_i$ and $j$ has remaining capacity, else 0). We set $\alpha=1$ and calibrate $\beta$ against empirical mobility data. The assignment is performed stochastically:
\begin{align}
j^* \sim \text{Categorical}(\{P_{ij}\}_j)
\end{align}
This probabilistic approach allows for a realistic distribution of commute distances, including long-tail commutes, unlike strict distance minimization.

\subsubsection{Derived Social Networks}
Multi-layer networks are a deterministic byproduct of the assigned home/school/work locations and institutional membership. While not used by the agent interface or the experiments in this paper, they are retained as an optional artifact for internal consistency checks and future extensions:
\begin{align}
G &= (V, E),\\
E &= E_{\text{household}} \cup E_{\text{home}} \cup E_{\text{school}} \cup E_{\text{work}} \nonumber\\
&\quad \cup E_{\text{neighborhood}} \label{eq:social_network}
\end{align}
where edges represent interaction opportunities induced by shared households, shared residential buildings, shared schools, shared workplace landuse, and neighborhood proximity.
To keep graphs sparse at scale, degrees are capped or edges are sampled within large buildings/institutions and edges can optionally be weighted by co-location frequency.

\subsubsection{Urban Environment Integration}
The platform integrates multiple layers of urban infrastructure:
\begin{align}
\mathcal{E} = \{\mathcal{P}, \mathcal{R}, \mathcal{B}, \mathcal{A}\} \label{eq:environment}
\end{align}
where:
\begin{itemize}
    \item $\mathcal{P}$: POI catalog with categorical attributes $\mathcal{P} = \{(p_i, \text{type}_i, \text{capacity}_i, \text{hours}_i)\}$
    \item $\mathcal{R}$: Road network graph $\mathcal{R} = (V_r, E_r, w_r)$ with edge weights (distance, speed, capacity)
    \item $\mathcal{B}$: Building set with spatial footprints and land use $\mathcal{B} = \{(b_i, \text{geom}_i, \text{use}_i, C_i)\}$
    \item $\mathcal{A}$: Administrative hierarchy (census blocks, districts, city) for spatial aggregation
\end{itemize}
When explicit capacities, opening hours, or road-capacity attributes are missing in the source layers, the implementation uses conservative defaults or simple rule-based proxies (e.g., POI-type-specific heuristics and road-class-based speed/capacity settings) to support feasibility checks.

\subsection{Activity Generation and Temporal Grounding}
\label{subsec:activity_generation}
We implement a \textbf{hybrid generative mechanism} to ensure both behavioral realism and temporal fidelity.
While the \textbf{sequence and semantics} of daily activities (e.g., the decision to visit a gym after work) are generated by the LLM-distilled policy to capture heterogeneous preferences, the \textbf{temporal attributes} (start time and duration) are grounded in the \textbf{National Time Use Survey}. Specifically, once an activity type is selected by the agent, its timing is sampled from the corresponding empirical distribution (e.g., 'Sports' duration distribution for a 'Gym' visit), thereby preventing unrealistic hallucinations common in pure LLM scheduling.

We utilize the \textbf{action initialization probability} (derived from activity start-time statistics) rather than the raw \textbf{action participation rate} (occupancy). Using raw occupancy rates as sampling probabilities—a common pitfall—would incorrectly bias the duration of activities. Our pipeline explicitly separates the \textit{decision to start} an activity from the \textit{duration} of the activity, ensuring that the generated temporal dynamics mathematically align with the aggregate census observations.


\subsection{Population Distribution Validation}
 
 We validate our synthetic population against census data at the tract level to ensure demographic accuracy.
 
 \subsubsection{Census Data Validation}
 
 Our population synthesis method generates 196,608 individuals across 89,988 private households in Higashihiroshima, with 5,641 additional group-quarter records retained separately from private-home assignment. We validate the synthetic population against 2020 Japanese Census tabulations at census-tract granularity across multiple demographic dimensions.
 
 For household size statistics, the census reports \emph{general household} counts, while some tracts include non-household residents (e.g., dormitories or institutional facilities). We therefore evaluate household size distributions on tracts where total population equals general-household persons (see Appendix~\ref{app:supplementary} for details).
 
 \paragraph{Distributional Fit Metrics}
 We distinguish hard constraints from soft-fit metrics. Tract-level total population is constrained to match census totals exactly, yielding very close agreement with census totals by construction. We therefore emphasize distributional similarity for variables not enforced as exact constraints.
 
 After restricting census tabulations to the instantiated study area, we obtain 198 finest-resolution census units (HYOSYO=2/4). We evaluate demographic fit on 185 tracts; 13 census units with zero population and zero households (e.g., industrial parks) are excluded.
 Gender ratios are well matched (male ratio MAE $< 0.02$). Age distributions achieve mean L1 = 0.1229 (median 0.10, max 0.31), mean KS = 0.0299 (max 0.12), and mean JS = 0.0047 (max 0.02), with 95\% of tracts having L1 $<$ 0.20. Household size distributions achieve mean L1 = 0.0547, mean KS = 0.0269, and mean JS = 0.0075. Employment counts (15+) show high tract-level agreement (R$^2 > 0.99$). Occupation distributions achieve mean L1 = 0.1945, mean KS = 0.0972, and mean JS = 0.0382. The tight distribution of per-tract errors reflects the effectiveness of the IPF constraints.
 
 \subsubsection{Spatial Distribution Validation}
 
 Unlike TAZ-based methods that assign residents to abstract zones, our building-level approach assigns households to specific georeferenced buildings under tract-level and capacity constraints. Because building footprints and land-use labels may be incomplete in a small number of tracts (e.g., industrial parks), we report explicit assignment diagnostics rather than silently forcing fallback placements.
 
 In our reference instantiation, all 89,988 private census-target households are successfully assigned to residential buildings. The 5,641 unmapped household records are group-quarter records and are intentionally excluded from private-home placement rather than treated as failed residential assignments.
 
 \subsubsection{School Assignment Validation}
 School assignment uses building-level home locations. Elementary and junior-high students are assigned by school-district polygons with nearest-school fallback. High-school assignment is nearest-school based with limited randomness among candidates within a distance threshold, and university assignment uses a gravity-style stochastic choice with weights proportional to $1/d^2$.
 
 In our reference instantiation, 42,376 out of 42,584 in-scope students are assigned to a school (99.51\%). The remaining 208 in-scope records are flagged as \texttt{too\_far\_stay\_home} by the diagnostic pipeline rather than silently assigned through unconstrained fallback rules; 747 school-age group-quarter residents are treated as out of scope for private home-based school assignment.
 We report the assigned school enrollment distribution in Figure~\ref{fig:school_enrollment}.
 
 \subsection{Mobility Pattern Validation}
 
 We compare commuting statistics against anonymized mobile phone mobility data from Yahoo Japan Mobility (YJMob100K) \cite{yabe2024yjmob100k}. The dataset discretizes location pings into 500m $\times$ 500m grid cells and timestamps into 30-minute bins, with the metropolitan area undisclosed for privacy. For our case study, we extract a subregion consistent with the Higashihiroshima area by registering the released mesh grid via manual rigid alignment. The registration uses coastline landmarks and major terrain features as control points, with an estimated alignment error of $<$500m (one grid cell). A sensitivity analysis indicates that commute distance distributions are robust to registration errors within this range. The extraction workflow is documented in the repository at \path{data_prepare/mobility_validation/README.md}, with comparison scripts under \path{data_prepare/mobility_validation/} and \path{data_prepare/step3_assignment/work/}; the comparison is treated as a commuting-distance diagnostic rather than an OD-flow benchmark.

\begin{figure}[htbp]
 \centering
 \includegraphics[width=\columnwidth]{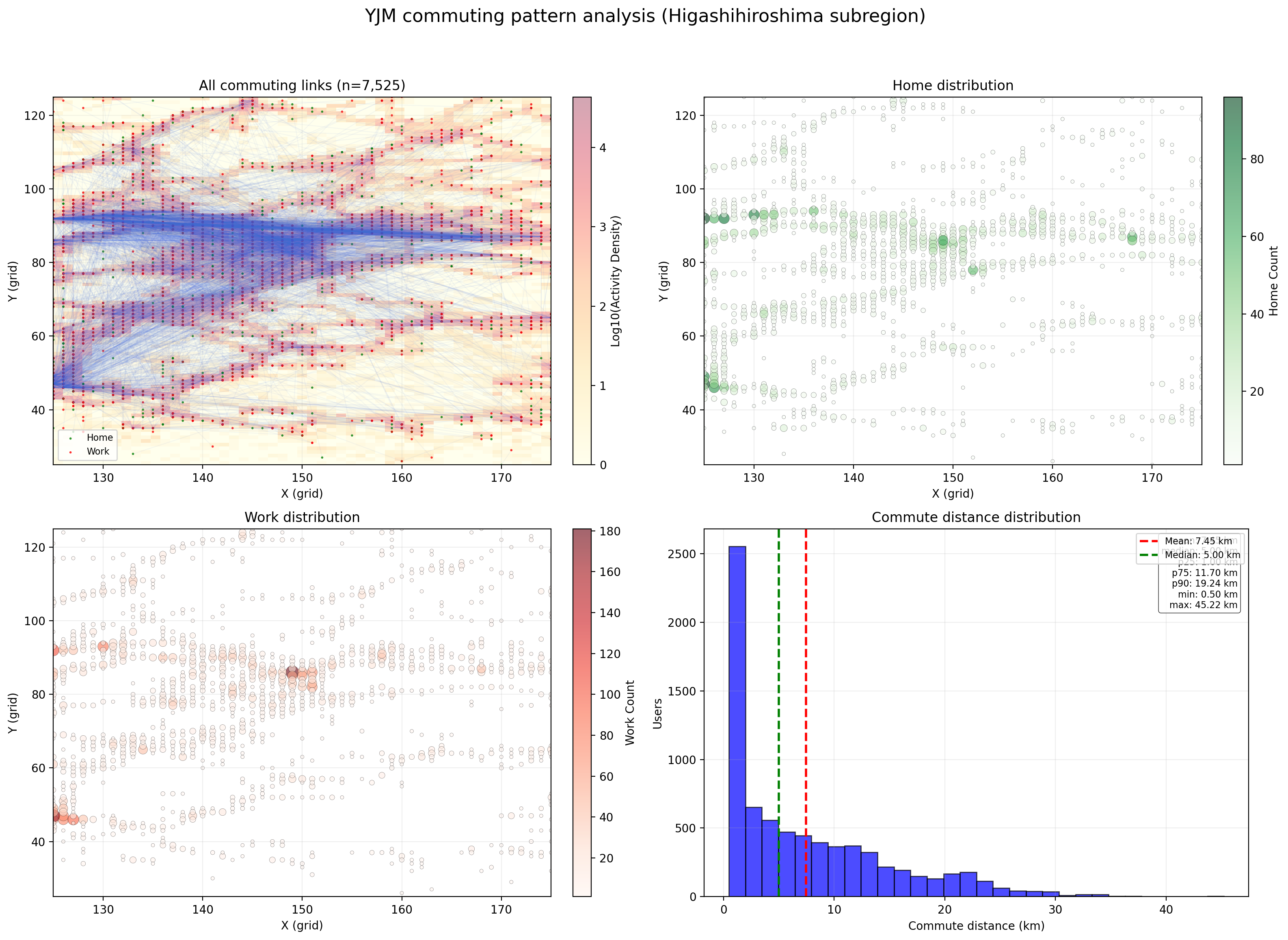}
 \caption{Commuting pattern extraction from YJMob100K after registering the anonymized mesh grid to our study area. The figure visualizes inferred home/work points and commuting distance statistics for the extracted subregion.}
 \label{fig:yjm_validation}
 \end{figure}

 We infer each user's home mesh from nighttime records and work mesh from daytime records (fixed time windows), then derive a commuting distance distribution in the mesh space. Figure~\ref{fig:yjm_validation} summarizes the extracted commuting patterns for the selected subregion.

 The released validation artifact extracts 7,525 YJM-derived commuters from the selected subregion, with a mean commute distance of 7.45~km, median of 5.00~km, and 90th percentile of 19.24~km. Against this reference, the 90,744 synthetic workers assigned to workplaces have a mean commute distance of 10.81~km, median of 10.09~km, and 90th percentile of 19.15~km. The resulting KS distance is 0.359 (0.399 when restricted to commutes $\leq$20~km). This diagnostic indicates that the current workplace assignment captures the upper-tail scale but underrepresents very short commutes and overrepresents 5--15~km trips. We therefore use YJM as a commuting-distance diagnostic and not as evidence of calibrated OD-flow prediction.

 Figure~\ref{fig:commute_distance} compares the resulting distributions. We treat this comparison as a commuting-distance diagnostic rather than a strict OD-flow correlation, because the observed mesh space is anonymized and requires manual registration.
 
 \begin{figure}[htbp]
 \centering
 \includegraphics[width=\columnwidth]{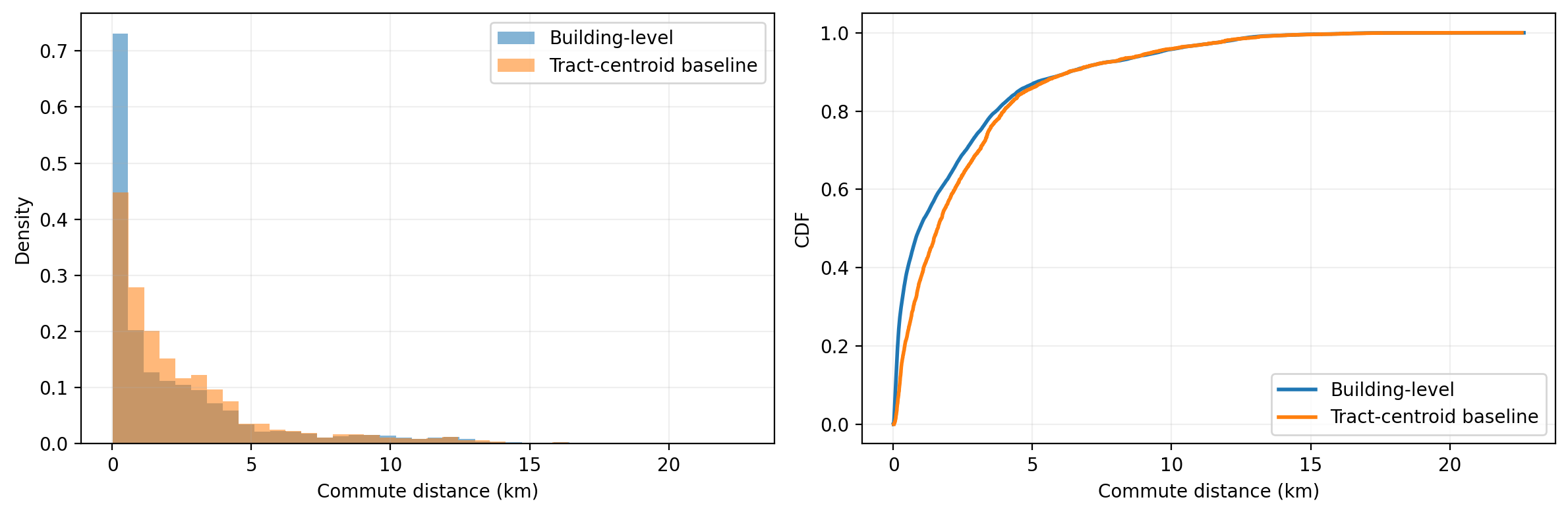}
 \caption{Commuting distance distributions under building-level grounding versus a tract-centroid baseline. The baseline collapses within-tract heterogeneity by placing all households at tract centroids, illustrating how coarse spatial grounding can distort short-range commuting structure even when workplace assignments are held fixed.}
 \label{fig:commute_distance_baseline}
 \end{figure}

\section{Platform Architecture}
\label{sec:architecture}

\begin{figure*}[t]
\centering
\includegraphics[width=0.75\textwidth]{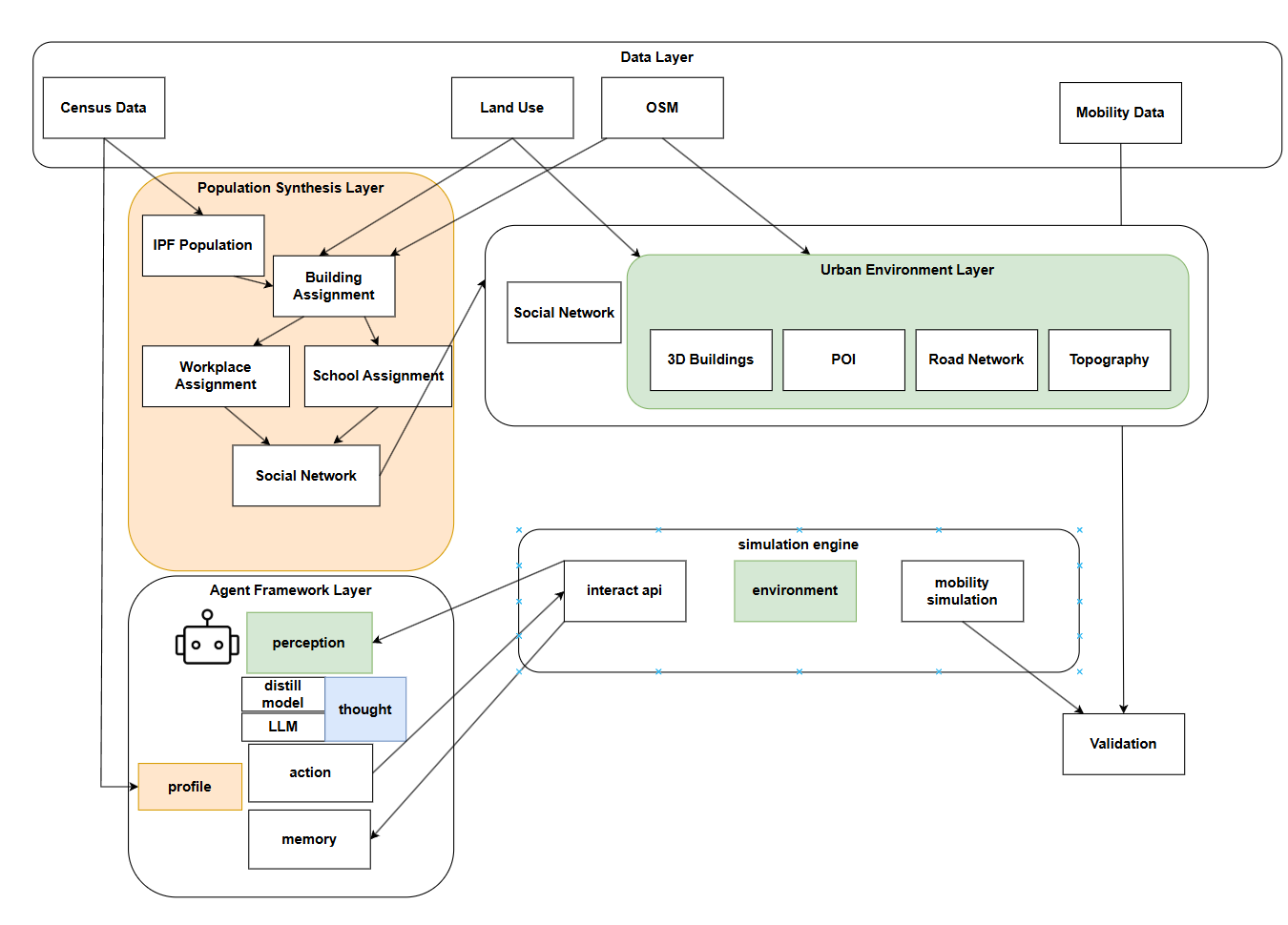}
\caption{GenWorld System Architecture. The platform is organized into three layers: Population \& Environment Foundation, Agent Decision Framework, and Simulation Engine. The architecture supports LLM integration and knowledge distillation for city-scale scalability.}
\label{fig:architecture}
\end{figure*}

GenWorld emphasizes \textbf{modularity} (independent components for flexibility), \textbf{scalability} (efficient handling of 200,000+ agents in our reference instantiation), and \textbf{accessibility} (LLM-compatible interfaces for AI researchers). Figure~\ref{fig:architecture} illustrates the detailed system architecture.
Platform UI screenshots (Streamlit-based interface) are provided in Appendix Figure~\ref{fig:platform_ui}.

\subsection{System Overview}

The platform is organized into three layers:

\paragraph{Layer 1: Population and Environment Foundation}
Instantiates the georeferenced urban world and synthetic population under census constraints and reports validation diagnostics; see Section~\ref{sec:grounding}.

\paragraph{Layer 2: Agent Decision Framework}
Exposes a structured agent--environment interface with binned observations and finite JSON-validated action candidates, enabling rule-based, teacher-LLM, and distilled-student policies; see Sections~\ref{sec:agent_interface} and~\ref{sec:distillation_scaling}.

\paragraph{Layer 3: Simulation Engine}
Orchestrates time-stepped multi-agent execution with feasibility checks, system-level consistency updates, and detailed logging; see Section~\ref{subsec:simulation_engine}.

The following subsections detail the simulation engine.
\subsection{Simulation Engine}
\label{subsec:simulation_engine}

The simulation engine orchestrates time-stepped multi-agent execution, managing time progression, spatial dynamics, and system-level feasibility constraints. The engine is designed to support both small-scale LLM experiments and large-scale distilled simulations.

\paragraph{Time-stepped Execution (Pseudo-code)}
The simulator advances in discrete time steps (typically 15-minute intervals) and executes validated actions under feasibility constraints, while recording structured decision traces for analysis and offline compilation.
\begin{algorithm}[htbp]
\caption{Time-stepped simulation engine with structured decision interface}
\label{alg:simulation_engine}
\begin{algorithmic}[1]
\For{each simulation step $t$}
  \State determine active agents $\mathcal{S}_t$ from schedules
  \For{each agent $i \in \mathcal{S}_t$}
    \State construct context $c_{i,t}$ from world state and persona
    \State $\tilde{o}_{i,t} \leftarrow \phi(c_{i,t}; q_t)$ \Comment{binned observation}
    \State $\mathcal{A}_{i,t} \leftarrow \kappa(q_t, \tilde{o}_{i,t})$ \Comment{finite candidates}
    \State $a_{i,t} \leftarrow \pi(\tilde{o}_{i,t}, \mathcal{A}_{i,t})$ \Comment{rule/teacher/student}
    \If{$v(\tilde{o}_{i,t}, a_{i,t}) = 0$}
      \State $a_{i,t} \leftarrow f(\tilde{o}_{i,t})$ \Comment{deterministic fallback}
    \EndIf
    \State execute $a_{i,t}$ and update agent/world states
    \State append decision record and trajectory log
  \EndFor
  \State apply system-level consistency updates (e.g., travel-time feedback and POI capacity)
  \State record aggregate metrics (e.g., utilization and travel-time indicators)
\EndFor
\end{algorithmic}
\end{algorithm}

This modular architecture supports repeatability through deterministic execution and configuration-based parameters, while enabling extensibility for new agent models, additional cities, and integration with external frameworks.

\section{Evaluation Cases and Scalability}
\label{sec:results}

\subsection{Evaluation Cases}

We report three simulation cases selected according to three criteria: (i) the case is generated by the public pipeline or the paper-figure reproduction contract, (ii) it has quantitative checks for schedule completeness and spatial feasibility, and (iii) it supports a specific claim about GenWorld rather than only serving as a visual showcase. Table~\ref{tab:evaluation_cases} summarizes the cases. In all reported runs, each person has a complete 1,440-minute daily schedule with no gaps or overlaps, and the activity-to-land-use compatibility checker reports zero violations.

\begin{table*}[t]
\centering
\caption{Evaluation cases included in the paper. Duration shares are computed over total person-minutes. Movement statistics use nonzero movement records between consecutive activity locations.}
\label{tab:evaluation_cases}
\small
\begin{tabular}{p{0.17\textwidth}p{0.14\textwidth}p{0.32\textwidth}p{0.21\textwidth}}
\toprule
\textbf{Case} & \textbf{Scope} & \textbf{Key evidence} & \textbf{Supported claim} \\
\midrule
Full-city weekday baseline & 196,608 agents, weekday normal & Home 79.25\%, duty 16.22\%, non-home 20.75\%; movement p95 = 17.26 km; no trips over 50 km & City-scale rollout over an empirically grounded synthetic population \\
Weekday--weekend contrast & 1,000-agent paired diagnostic & Duty decreases from 15.68\% on weekday to 0\% on weekend; leisure increases from 3.81\% to 37.92\%; no land-use violations & The same population can express different temporal regimes under controlled day-type constraints \\
Warning-response perturbation & 1,000 agents, weekday alarm at 15:00 & After the warning, all agents are at home at 15:00, 16:00, and 18:00; emergency-return-home occupies 37.5\% of person-minutes & Scenario perturbations can trigger schedule replanning and produce auditable response traces \\
\bottomrule
\end{tabular}
\end{table*}

\subsubsection{Case 1: Full-City Weekday Baseline}

The full-city baseline simulates 196,608 agents distributed across 89,988 private households, with additional group-quarter records retained separately from private-home assignment. Building-level home assignment, home/school/work anchors, and daily activity schedules are executed under the structured interface. The run produces 947,233 activity records. Each agent receives a complete daily schedule from midnight to midnight. Aggregated by duration, home activities account for 79.25\% of person-minutes, duty activities (work and study) for 16.22\%, leisure for 2.43\%, and maintenance activities for 2.11\%. The median nonzero movement distance is 5.59 km, the 95th percentile is 17.26 km, and no movement exceeds 50 km.

\begin{figure}[!htbp]
\centering
\begin{subfigure}[b]{\columnwidth}
    \centering
    \includegraphics[width=\textwidth]{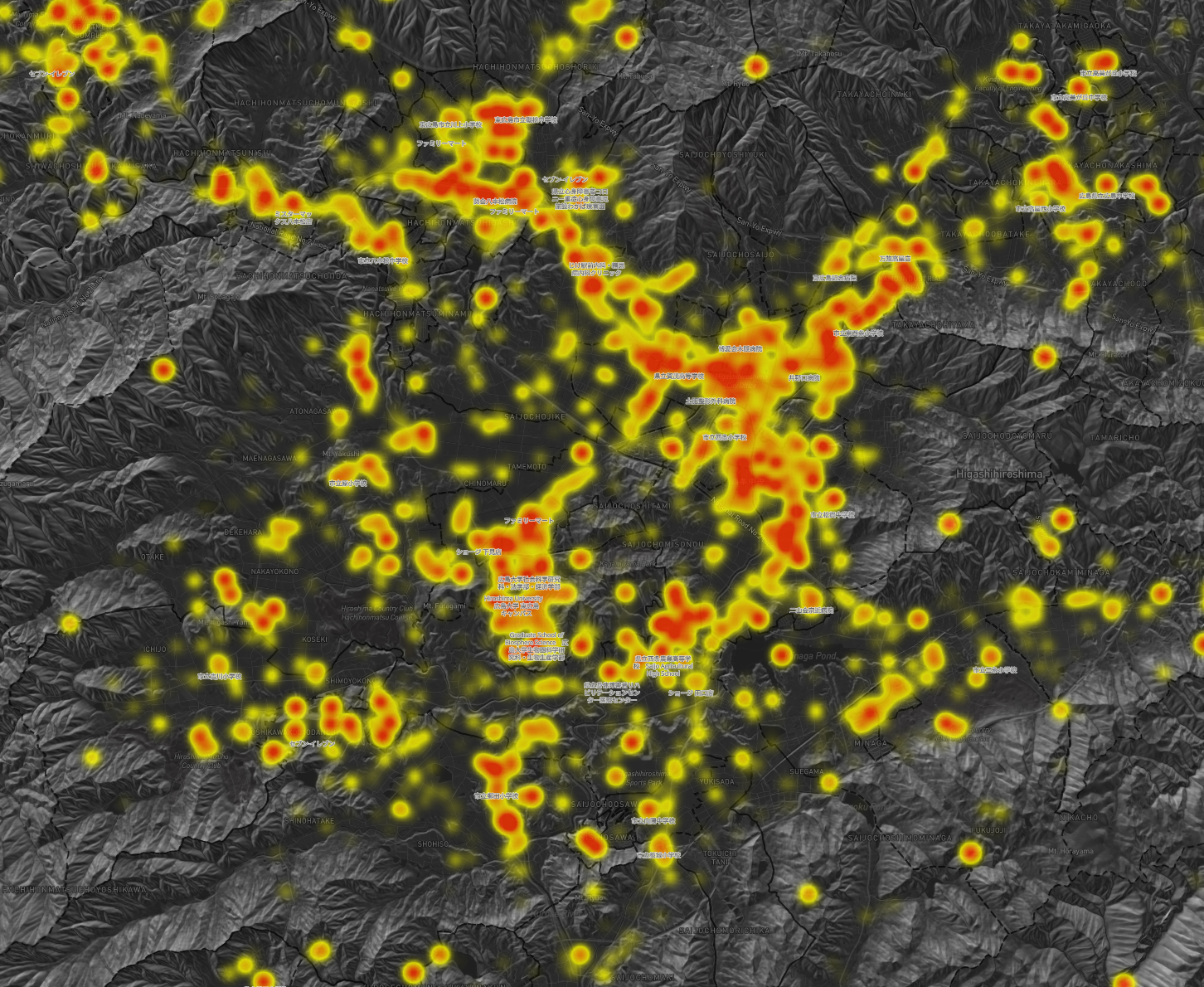}
    \caption{Worktime resident-location heatmap.}
\end{subfigure}

\begin{subfigure}[b]{\columnwidth}
    \centering
    \includegraphics[width=\textwidth]{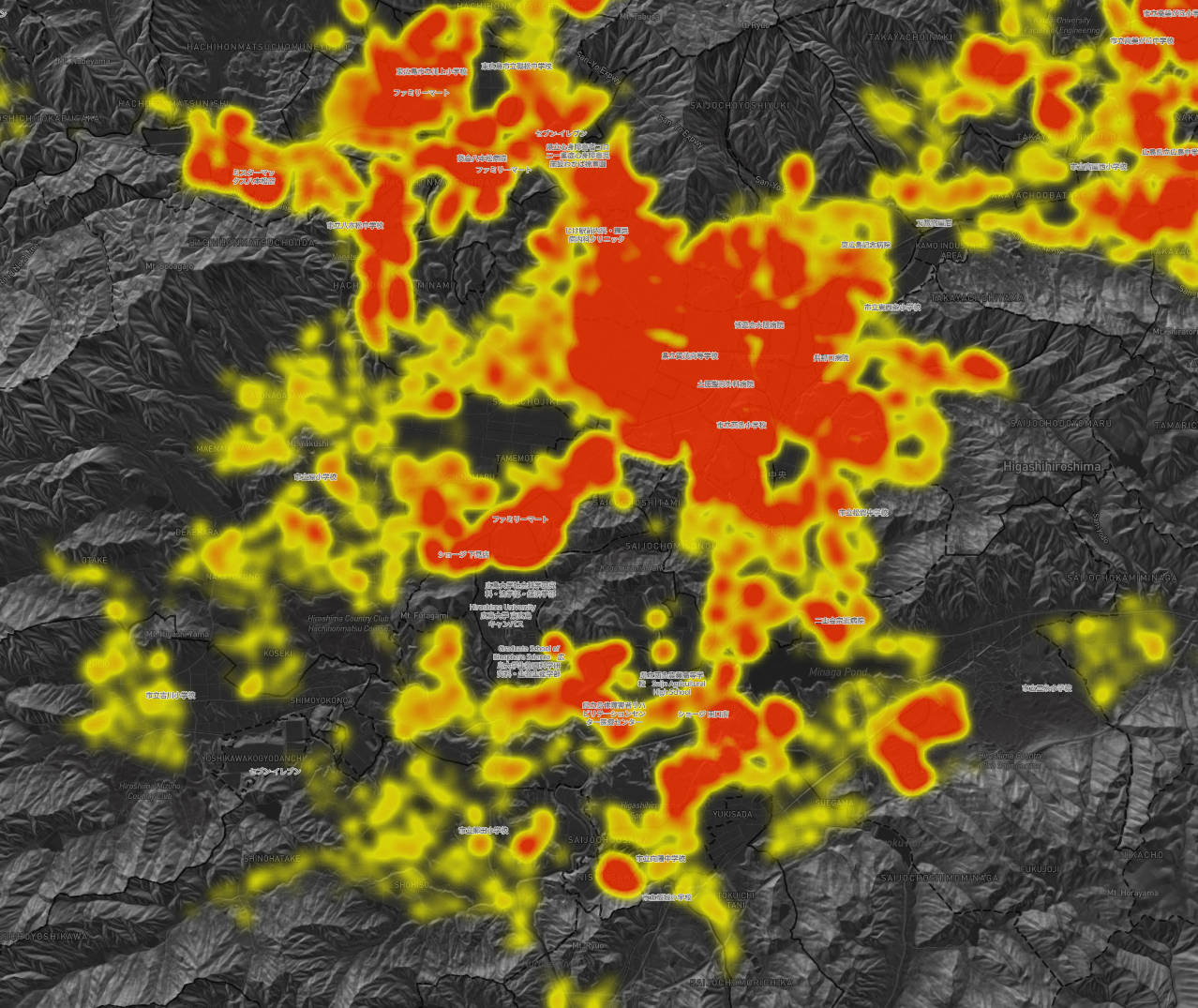}
    \caption{Nighttime resident-location heatmap.}
\end{subfigure}
\caption{Day--night contrast of visualized resident locations in the baseline rollout. The worktime snapshot highlights dense daytime clustering around major institutional and employment centers (e.g., the Hiroshima University area), whereas the nighttime snapshot shows these areas becoming nearly empty as the population shifts back toward residential neighborhoods.}
\label{fig:day_night_population}
\end{figure}

We visualize the spatial distribution of agents and their daily commuting flows. The 3D visualization supports qualitative inspection of residential density gradients, commuting corridors, activity hotspots around commercial and institutional areas, and day--night population shifts. Figure~\ref{fig:day_night_population} shows two snapshots of the visualized resident locations: during worktime the distribution exhibits strong clustering around activity centers (e.g., the Hiroshima University area), while at nighttime these daytime hotspots become sparse as residents return to their home neighborhoods.

Additional weekday spatial heatmaps for representative activity types (shopping, socializing, and childcare) at multiple time windows are provided in the appendix (Figure~\ref{fig:activity_time_spatial_3x5}).

\begin{figure}[!htbp]
\centering
\includegraphics[width=\columnwidth]{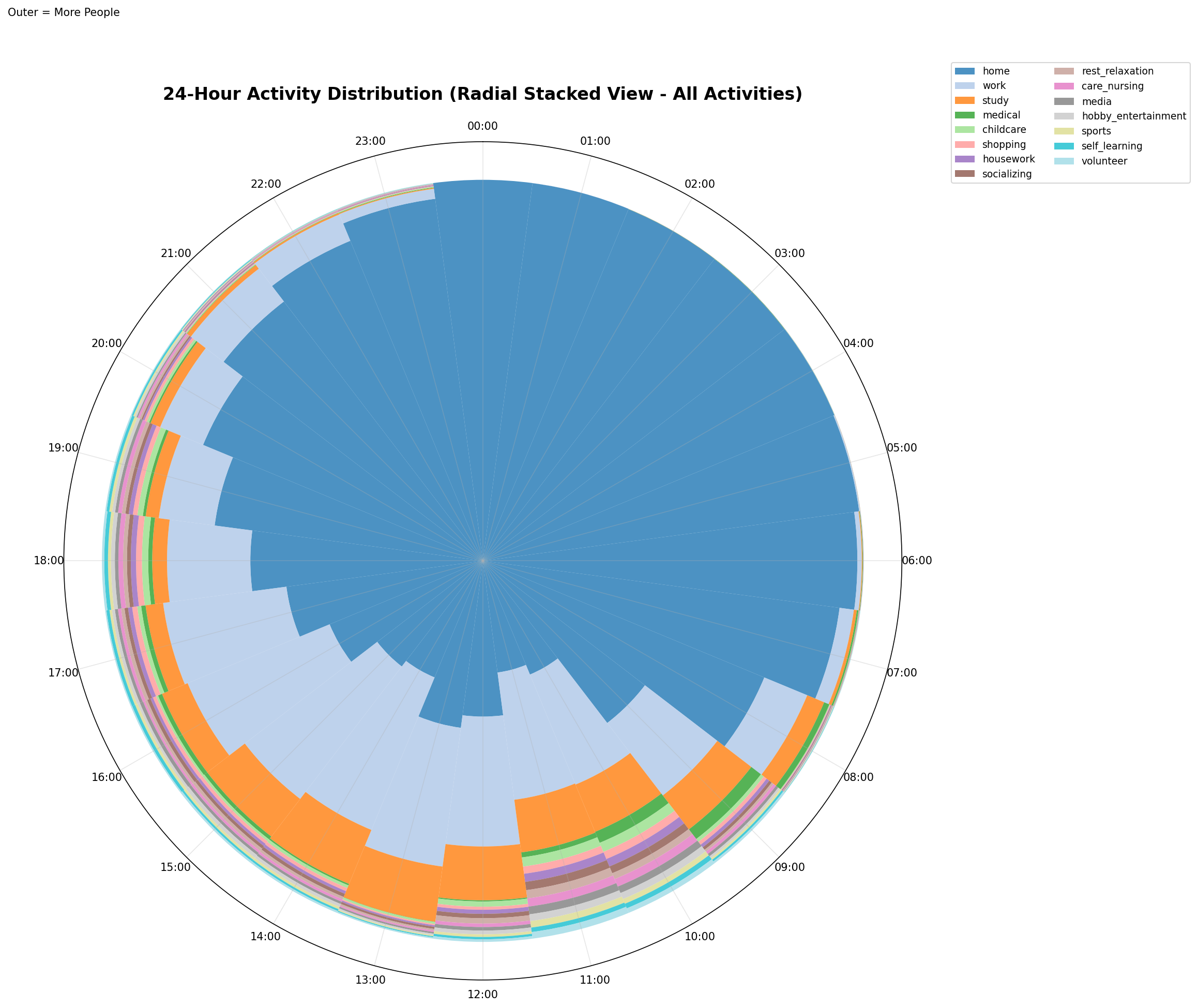}
\caption{24-hour activity occupancy distribution in the baseline rollout, shown as a radial stacked plot (outer radius indicates more people). The visualization highlights the expected day--night cycle: home/sleep dominates overnight, work and study increase during daytime hours, and leisure and other discretionary activities rise in the evening.}
\label{fig:activity_24h_radial}
\end{figure}

We also summarize the city-scale diurnal rhythm by aggregating simulated activity occupancy over time. Figure~\ref{fig:activity_24h_radial} visualizes the 24-hour distribution of activity categories as a radial stacked plot, providing a compact view of time-of-day regularities in the baseline rollout.

\subsubsection{Case 2: Weekday--Weekend Behavioral Contrast}

To test whether the same population foundation can support different temporal regimes, we compare two 1,000-agent diagnostic runs under weekday and weekend settings. Both runs pass the same schedule-completeness and land-use compatibility checks. The weekday run contains 5,016 records and includes work/study duty activities, while the weekend run contains 5,000 records and removes duty activities by construction.

The contrast is clear at the duration level. In the weekday run, duty activities account for 15.68\% of person-minutes, while leisure activities account for 3.81\%. In the weekend run, duty falls to 0\% and leisure rises to 37.92\%. The spatial pattern also changes: the weekend movement-distance distribution is shorter (p95 = 5.67 km) than the weekday distribution (p95 = 13.85 km), reflecting more local discretionary activity in the diagnostic sample. These results do not validate weekend behavior against independent observations; instead, they show that GenWorld can apply different day-type constraints to the same synthetic population while preserving schedule and land-use consistency.

\subsubsection{Case 3: Warning-Response Perturbation}

We include a warning-response case as a controlled perturbation test rather than as a calibrated disaster model. Starting from a weekday setting, an alarm is introduced at 15:00. The policy then replans subsequent activity segments under a rule-constrained emergency response, producing explicit \texttt{emergency\_return\_home} records and preserving the same schedule-completeness constraints.

In the 1,000-agent alarm run, the simulator produces 6,000 activity records. The run passes the land-use compatibility checker with zero violations. At 15:00, 16:00, and 18:00, all 1,000 agents are at home according to the diagnostic evaluator. Emergency-return-home records account for 37.5\% of total person-minutes. This case demonstrates that GenWorld can inject a scenario perturbation, replan schedules, and produce auditable response traces. It should be interpreted as an illustrative warning-response stress test, not as evidence that the current implementation predicts real evacuation behavior or fully implements a psychological theory of disaster response.

\subsubsection{Road-Flow Visualization}

\begin{figure}[!htbp]
\centering
\includegraphics[width=\columnwidth]{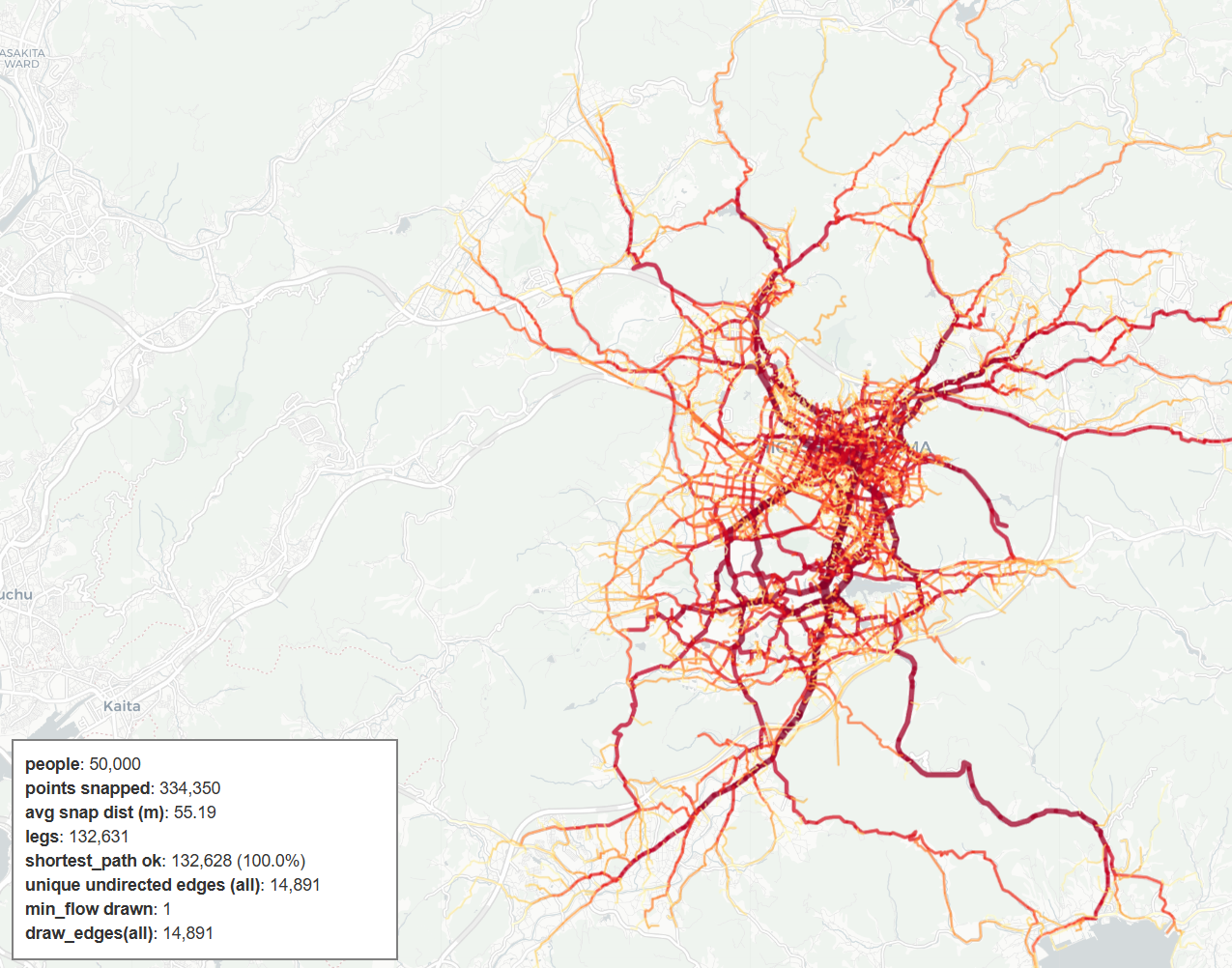}
\caption{All-day road-network traffic flow aggregated from a 50,000-resident sample. Trips are routed via static shortest paths (no congestion feedback); edge intensity indicates accumulated volume. This is intended as a visualization of spatial demand patterns rather than a validated traffic simulation.}
\label{fig:road_flow_allday}
\end{figure}

We further visualize aggregate road-network traffic flow by routing simulated trips between consecutive activity locations. Figure~\ref{fig:road_flow_allday} shows the all-day flow map computed from a 50,000-resident sample, where edge color intensity indicates higher accumulated volumes. Note that this is a \emph{static shortest-path visualization} without dynamic congestion feedback; validating against real-time traffic counts and incorporating equilibrium assignment are left for future work.

\subsubsection{Scalability Analysis}

Through offline compilation, simulation-time decision-making can be implemented as amortized constant-time table lookup and sampling under bounded candidate sets. The computational complexity comparison is as follows:

\begin{itemize}
\item \textbf{Online LLM}: $O(N \cdot T \cdot C_{\text{LLM}})$ per simulated day, where $N$ is agent count, $T$ is decision steps per day, and $C_{\text{LLM}}$ is per-query LLM inference cost (typically 0.5--2s for local 7B models).
\item \textbf{Distilled policy}: $O(N \cdot T \cdot C_{\text{lookup}})$, where $C_{\text{lookup}} \approx 1\mu$s (hash table lookup + categorical sampling).
\end{itemize}

For $N = 200{,}000$ agents with $T = 96$ decision points per day (15-minute steps), online LLM simulation would require $\sim$19M inference calls per simulated day, which is computationally expensive in practice. Our distilled policy replaces these calls with table lookups, allowing city-scale rollout in our reference setup.

In a micro-test, Python lookup achieves 1.85M queries/s (0.54$\mu$s per query) over 200,000 randomized context keys on an Intel Core i5-14600K CPU. End-to-end wall-clock time per simulator step also includes environment updates, spatial queries, and activity execution; profiling under varying agent counts is ongoing work.

\subsection{Summary}

These cases show three aspects of GenWorld: city-scale rollout over an empirically grounded population, controlled behavioral contrast across day types, and auditable replanning under a warning perturbation. The results support GenWorld as a reproducible simulation infrastructure. They do not by themselves establish calibrated forecasting performance for traffic, evacuation, or policy outcomes; broader validation would require additional external observations and scenario-specific calibration.

\section{Discussion}
\label{sec:discussion}

\paragraph{Limitations and Future Work}
Several limitations remain in the current reference instantiation.

\paragraph{Validation Scope}

We validate synthetic populations against census tabulations, commuting distances against YJMob100K mobile phone data, and activity schedules against the Japanese National Time Use Survey (e-Stat). Our activity schedule validation shows good agreement for diurnal patterns (average correlation $r > 0.86$, RMSE $< 3\%$), though peak-time shifts for work/study activities suggest lunch-break modeling needs refinement. Broader validation, such as link-level traffic counts and full OD-flow correlation, would require additional calibrated datasets and is left for future work.

\paragraph{Distillation Fidelity}

Our distillation pipeline aggregates teacher-model responses into lookup tables, but the fidelity of this compilation is not fully validated. We use $K=10$--$30$ samples per context key with a single teacher model (Gemma 3 27B); ablation of sampling count, temperature, and teacher model choice is needed. We also do not quantitatively compare distilled outputs against fresh teacher queries (e.g., via KL divergence or decision agreement rate).

\paragraph{Behavioral Modeling}

The structured interface enables logging and analysis of LLM-driven decisions, but connecting these to human decision processes is not addressed here. Possible extensions include comparisons against human subjects or stated-preference surveys, sensitivity analyses of prompt design, and evaluation of emergent behaviors under scenario perturbations.

\paragraph{Case-Study Boundaries}

The three cases in Section~\ref{sec:results} are intended to demonstrate infrastructure capabilities under controlled settings. The full-city weekday case supports scalability and schedule-generation claims, while the weekday--weekend comparison demonstrates that the same population can be simulated under different temporal constraints. The alarm case is more limited: it is a warning-response perturbation with rule-constrained replanning and should not be interpreted as a calibrated evacuation model, a validated disaster-response forecast, or a full implementation of Protection Motivation Theory. Its role is to show that GenWorld can inject a scenario event and produce auditable response traces for later behavioral calibration.

\paragraph{Generalizability}

The current implementation is instantiated in Higashihiroshima, a mid-sized Japanese city with approximately 200,000 residents. Higashihiroshima has a relatively dispersed urban form centered around Hiroshima University; scalability to denser metropolitan areas (Tokyo, Osaka) with more complex transit networks remains untested, and computational challenges may arise at 10$\times$ population scales.

Our data pipeline relies on Japan-specific sources (e-Stat census, YJMob100K mobility, Hiroshima DoBOX land use). Replication elsewhere requires equivalent data sources and adapted preprocessing; availability and format consistency vary across regions. Activity patterns and commuting behaviors also differ across urban contexts---US suburban sprawl, European compact cities, and Asian high-density development each have distinct characteristics. The distilled decision distributions may not transfer without local calibration.

\paragraph{Potential Application Scenarios}
Although the results reported in this paper focus on empirical grounding, scalable rollout, and controlled scenario diagnostics, the same instantiation and structured decision traces can support qualitative what-if analyses. Example extensions include transportation-demand inspection under hypothetical transit or land-use changes, warning-response experiments with richer behavioral models, and urban policy diagnostics under routine or capacity modifications. Such uses require scenario-specific assumptions, calibration data, and validation metrics before they can be treated as forecasts or decision-support evidence.

\section{Conclusion}
\label{sec:conclusion}

This paper introduced GenWorld as an empirically grounded urban simulation infrastructure for scalable LLM-agent studies. The central problem is not population synthesis alone or LLM distillation alone, but the connection between the two: LLM-agent simulations need realistic urban constraints, while city-scale rollout cannot rely on online LLM calls for every agent decision.

GenWorld addresses this grounding--scaling gap through a connected system design. A building-level synthetic urban world provides census-consistent population structure, spatial anchors, and land-use constraints. A structured agent interface maps city and persona states into binned observations, finite candidate sets, JSON-valid actions, deterministic execution semantics, and machine-readable traces. Offline policy compilation then shifts repeated teacher-model queries out of the rollout loop and executes compiled stochastic policies through lightweight lookup and sampling.

The Higashihiroshima instantiation demonstrates the feasibility of this infrastructure for 196,608 synthetic residents, with demographic validation against census tabulations, commuting-distance diagnostics against YJMob100K, and reproducible evaluation cases covering full-city weekday rollout, weekday--weekend contrast, and warning-response perturbation. These results support GenWorld as a platform for grounded and scalable LLM-agent experimentation. They do not establish calibrated forecasting performance for transportation, evacuation, or policy analysis; such applications require additional behavioral calibration, external validation data, and scenario-specific evaluation metrics. Code, configurations, documentation, and a deterministic public demo are available in the project repository, following the principles of reproducible urban research~\cite{felix2025reproducible}.

\section*{Code and Data Availability}
Code, configuration files, documentation, and a deterministic public demo are available at \url{https://github.com/Perseus1993/genworld}. The repository includes the staged data-preparation pipeline, the public demo for tract 34212058004, and a paper-figure reproduction manifest. Large source datasets and generated outputs are not redistributed in the repository. Open or registration-based inputs should be obtained from their original providers and placed according to the paths documented in the repository. YJMob100K-derived inputs are non-redistributable and are used only as local validation inputs. The arXiv source package contains only the manuscript sources and figures needed to reproduce the paper PDF.

\ifblind\else
\section*{Acknowledgments}
We thank Xuesong (Simon) Zhou for his valuable suggestions.
\fi

\bibliographystyle{plain}
\bibliography{references}

\appendix
\section{Supplementary Materials}
\label{app:supplementary}

\setcounter{figure}{0}
\renewcommand{\thefigure}{A\arabic{figure}}

\subsection{Additional Figures}

\begin{figure}[htbp]
\centering
\includegraphics[width=\columnwidth]{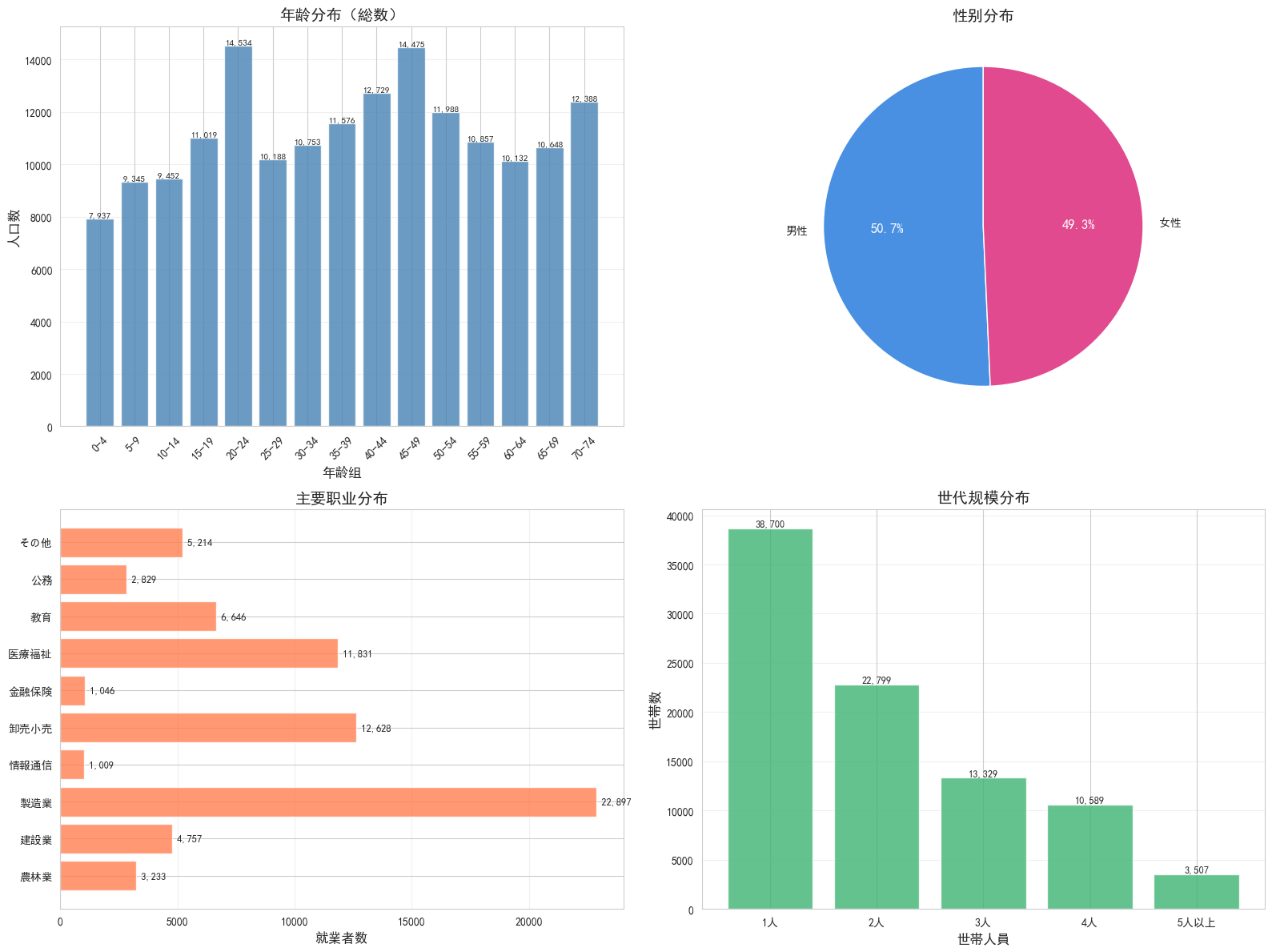}
\caption{Census data summary showing age-gender-occupation distributions across the finest-resolution census units (level 2 + level 4) in Higashihiroshima. The tabulations are used as a reference for evaluating demographic accuracy of the synthetic population.}
\label{fig:census_data}
\end{figure}

\begin{figure}[htbp]
\centering
\includegraphics[width=\columnwidth]{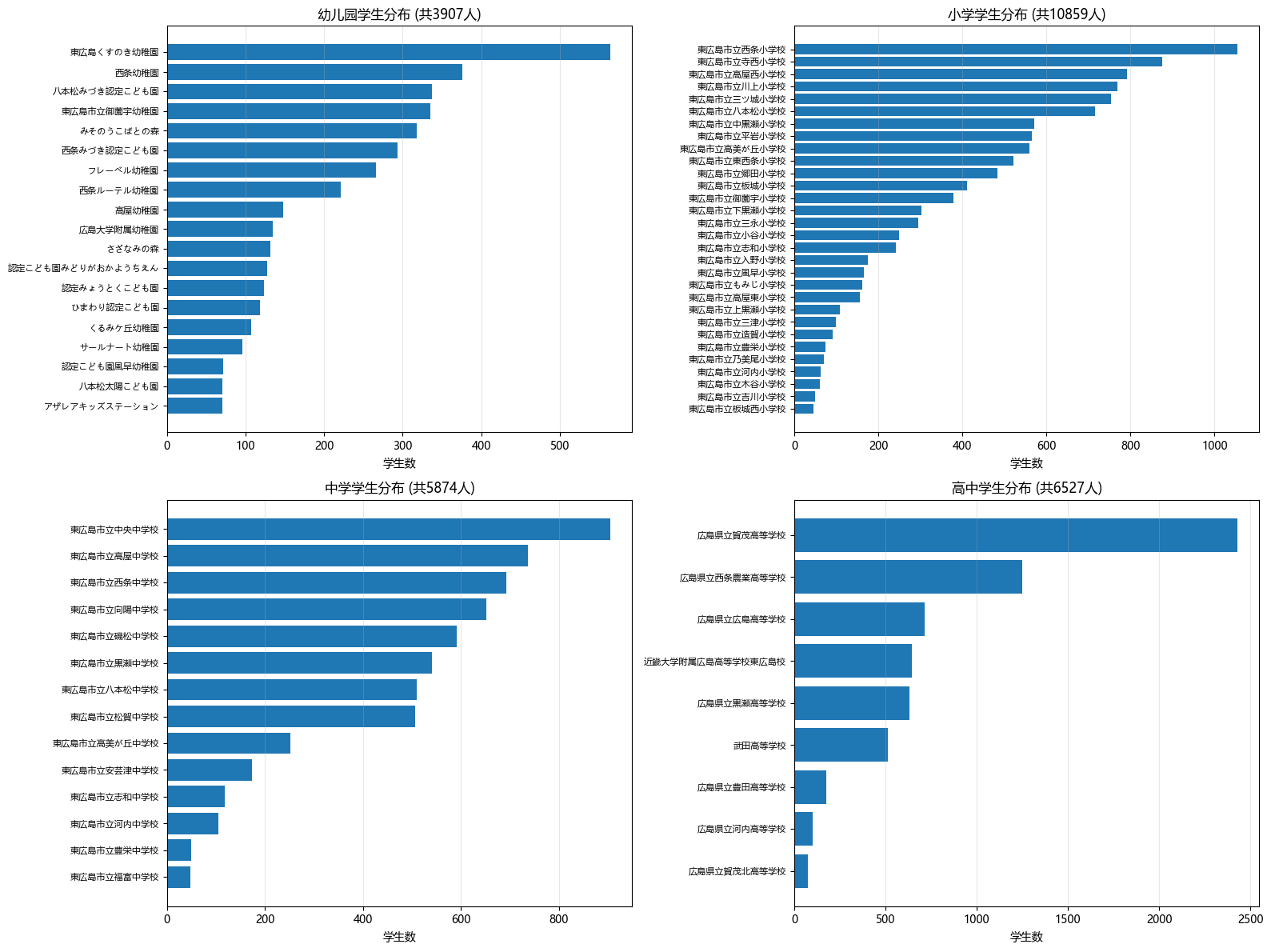}
\caption{School enrollment distribution across 85 schools in Higashihiroshima, showing the number of students assigned to each educational level. The distribution is consistent with official enrollment statistics.}
\label{fig:school_enrollment}
\end{figure}

\begin{figure}[htbp]
\centering
\includegraphics[width=0.95\columnwidth]{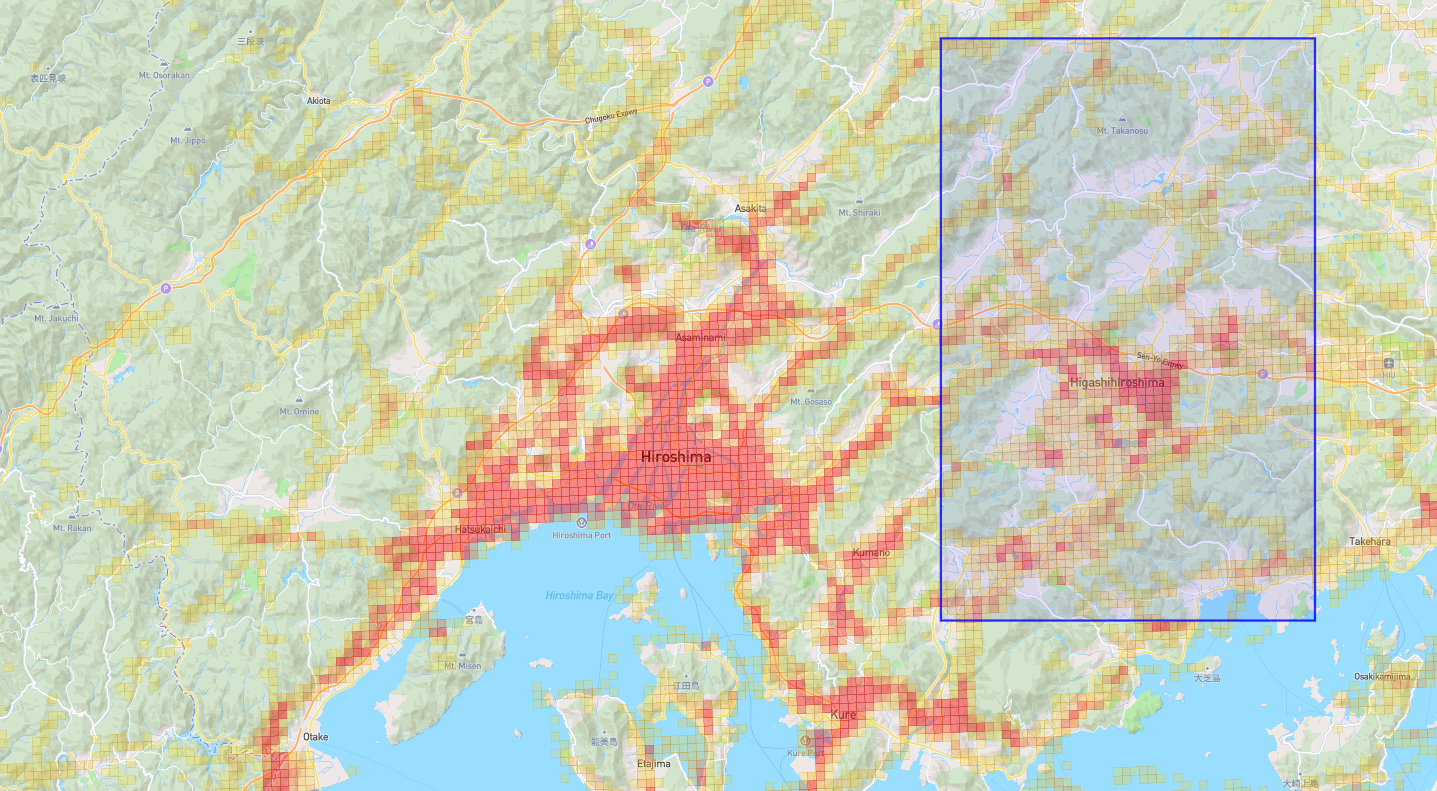}
\caption{Example of YJMob100K data showing aggregated commuting flows after registering the anonymized mesh grid to our Higashihiroshima study area. The data provides mesh-level origin-destination patterns derived from anonymized mobile phone GPS trajectories, and is used as an external mobility reference.}
\label{fig:yjm_example}
\end{figure}

\begin{figure}[htbp]
\centering
\includegraphics[width=\columnwidth]{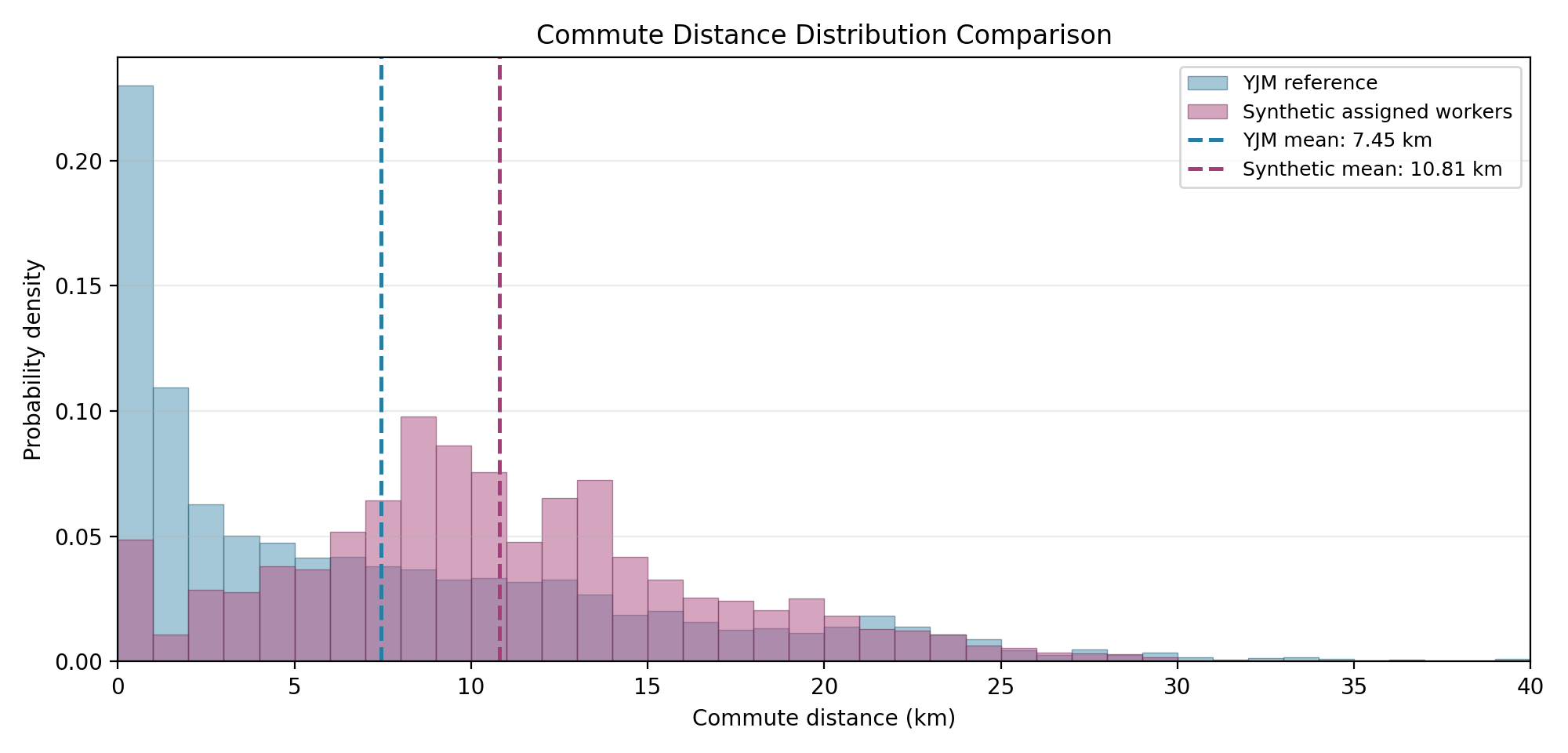}
\caption{Commute distance distribution comparison between the synthetic population and YJM data, used as a diagnostic for commuting-distance scale.}
\label{fig:commute_distance}
\end{figure}

\begin{figure*}[htbp]
\centering
\begin{subfigure}[b]{0.44\textwidth}
    \centering
    \includegraphics[width=0.95\textwidth]{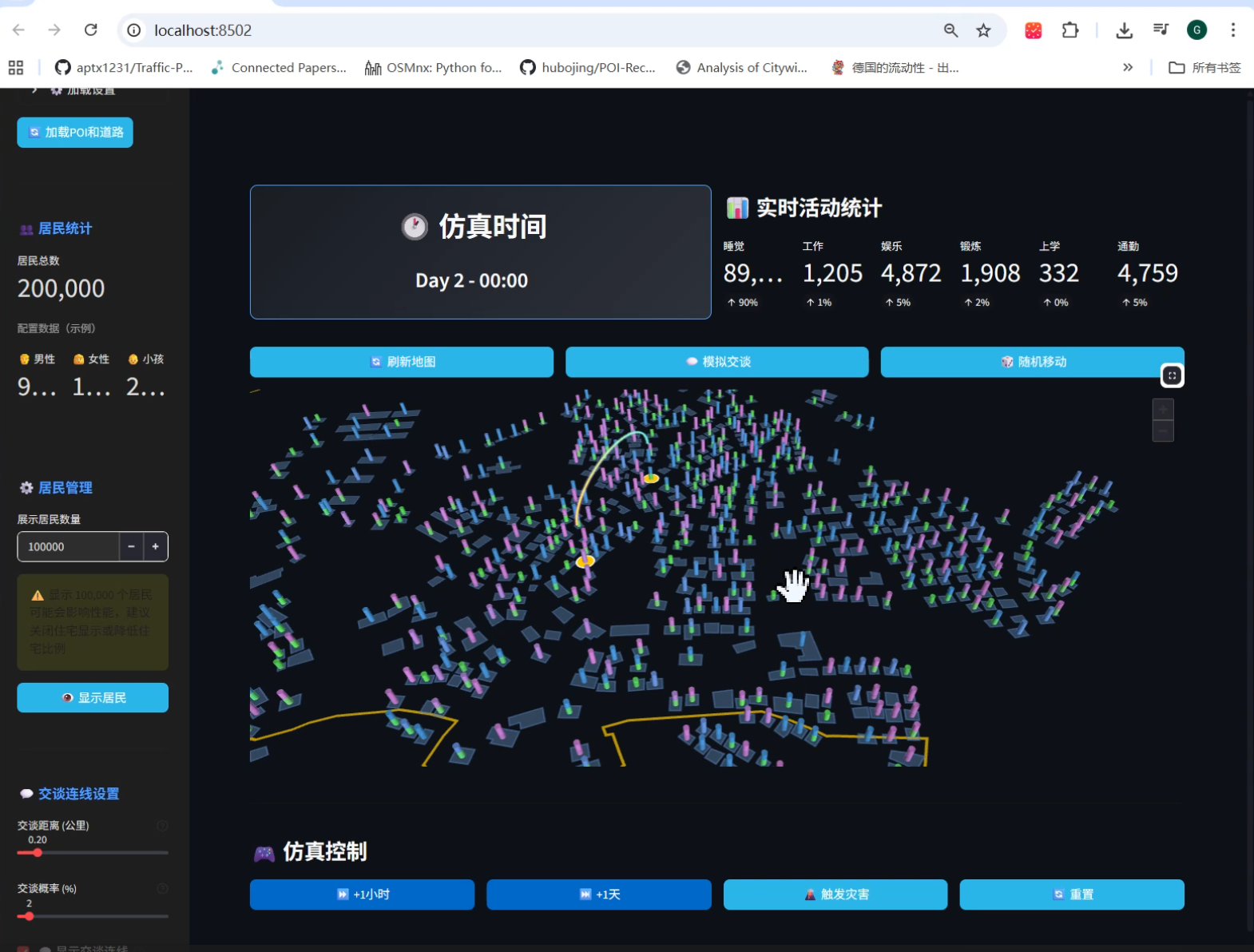}
    \caption{Simulation dashboard and real-time activity statistics in the Streamlit-based UI.}
\end{subfigure}
\hfill
\begin{subfigure}[b]{0.44\textwidth}
    \centering
    \includegraphics[width=0.95\textwidth]{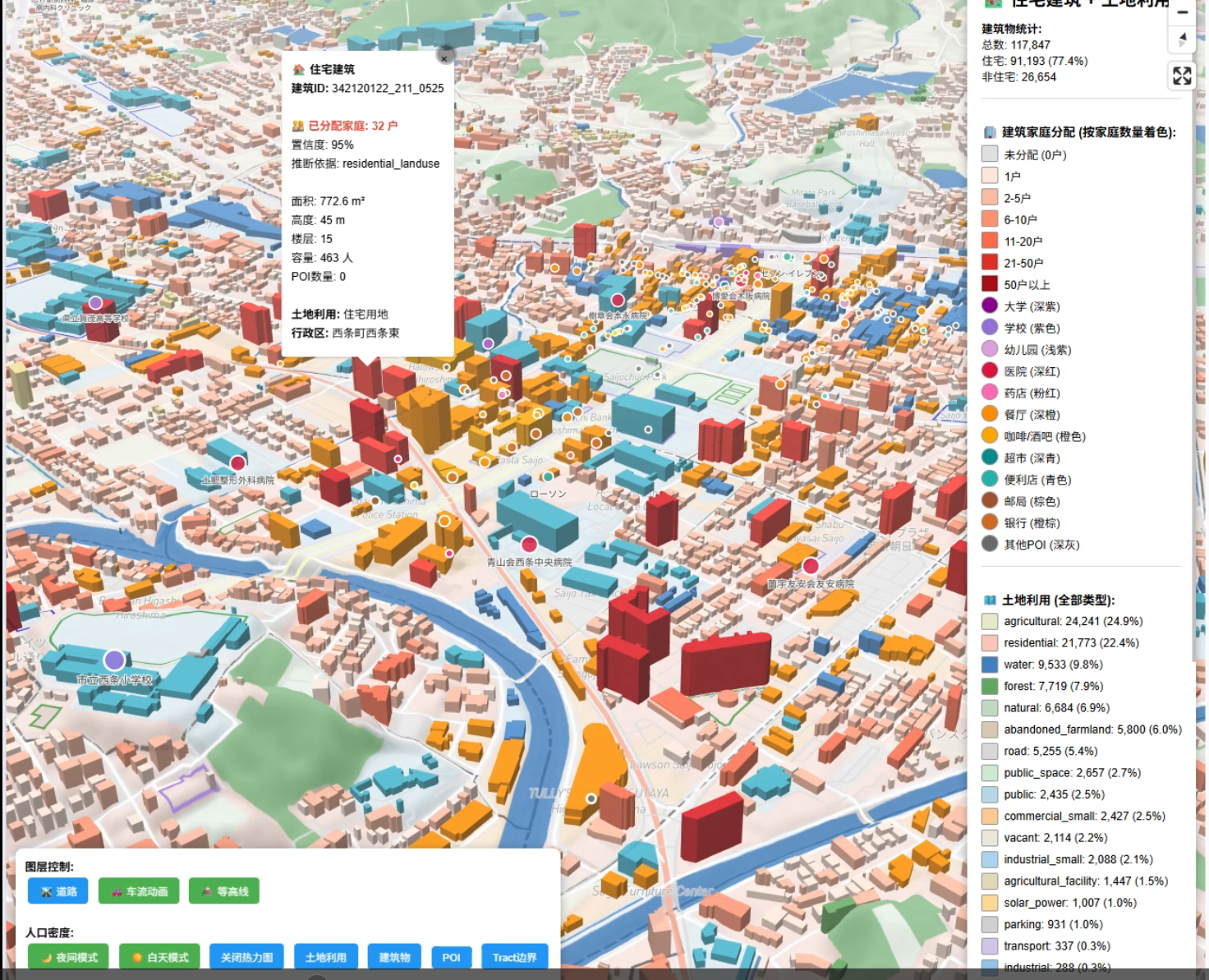}
    \caption{Interactive building-level map view for inspecting the instantiated urban world (e.g., land use and assigned households). Residential buildings are rendered in red with color intensity proportional to resident counts (darker indicates more residents).}
\end{subfigure}
\caption{Platform UI screenshots of GenWorld, implemented with Streamlit for interactive inspection and monitoring of the simulation and instantiated urban world.}
\label{fig:platform_ui}
\end{figure*}

\begin{figure*}[!htbp]
\centering
\includegraphics[width=0.95\textwidth]{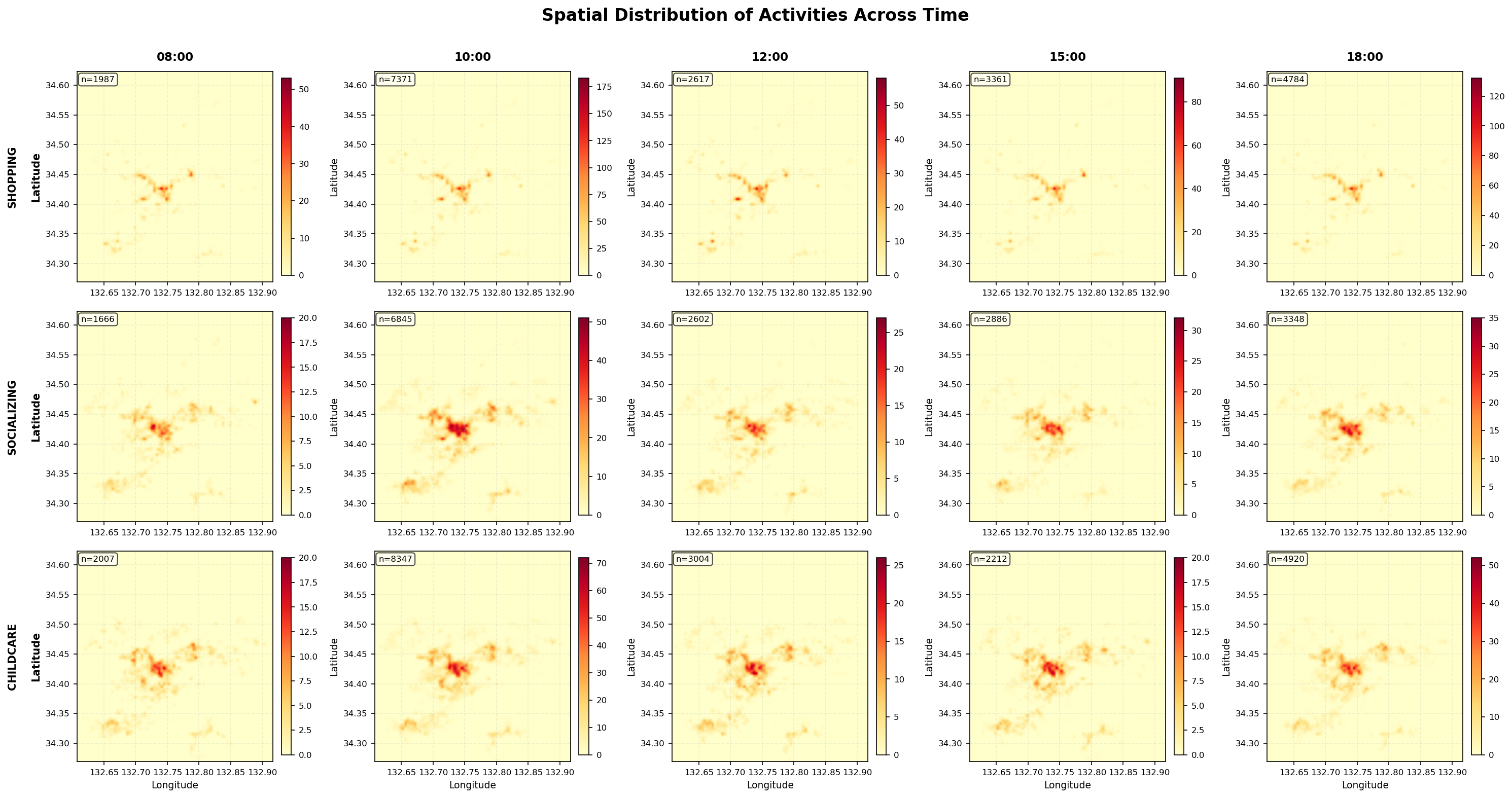}
\caption{Weekday spatial heatmaps for three representative activity types (shopping, socializing, and childcare) at five time windows. Each row corresponds to an activity type and each column corresponds to a time window; color intensity indicates higher occupancy.}
\label{fig:activity_time_spatial_3x5}
\end{figure*}

\subsection{Data Sources}
\label{app:data_sources}

\begin{table*}[!htbp]
\centering
\small
\setlength{\tabcolsep}{4pt}
\renewcommand{\arraystretch}{1.05}
\caption{Data sources used to instantiate and validate GenWorld in Higashihiroshima. Access column indicates availability: \textbf{Open} = publicly available for automatic download; \textbf{Reg} = requires free registration; \textbf{NR} = non-redistributable (requires user to obtain from original source).}
\label{tab:data_sources}
\begin{tabular}{@{}p{0.15\textwidth} p{0.15\textwidth} p{0.08\textwidth} p{0.52\textwidth}@{}}
\toprule
\textbf{Data Type} & \textbf{Source} & \textbf{Access} & \textbf{Description}\\
\midrule
Census Data & \href{https://www.e-stat.go.jp/en}{e-Stat} & Open & Age-gender, household, occupation statistics (198 census units)\\
Time Use Survey & \href{https://www.e-stat.go.jp/en}{e-Stat} & Open & National time-use survey tabulations for activity distributions\\
Admin Boundaries & \href{https://www.e-stat.go.jp/en}{e-Stat} & Open & Census tract boundaries for spatial aggregation\\
Buildings & OpenStreetMap & Open & Building footprints with height and area (45,000+ buildings)\\
POI Data & OpenStreetMap & Open & Points of interest (57,000 POIs)\\
Manufacturing POIs & Hiroshima High-Tech Assoc. & NR & Company locations and employee counts (215 facilities)\\
Land Use & \href{https://hiroshima-dobox.jp/en}{Hiroshima DoBOX} & Reg & Parcel-level land use classification\\
Elevation & \href{https://fgd.gsi.go.jp/download/}{GSI FGD DEM1A} & Reg & 1m-mesh digital elevation model\\
Road Network & OpenStreetMap & Open & Road network with hierarchy (15,861 nodes)\\
School Districts & \href{https://www.e-stat.go.jp/en}{e-Stat} & Open & School district boundaries (85 schools)\\
Mobile Phone Data & YJMob100K~\cite{yabe2024yjmob100k} & NR & Aggregated commuting patterns for validation\\
\bottomrule
\end{tabular}
\end{table*}

\FloatBarrier
\subsection{Intention and Activity-Type Taxonomy}
\label{app:intention_activity_taxonomy}

\paragraph{Activity Type Vocabulary}
 We use a small, discrete activity-template vocabulary (configured in \path{data_prepare/step4_activity/bins_activity_preference.json}) in our reference instantiation:
 \begin{Verbatim}[fontsize=\footnotesize,breaklines=true,breakanywhere=true]
 sleep_rest, work_task, study_class, daily_shopping, personal_service, solo_meal, social_meal, medical_care, admin_errand, social_visit, entertainment_activity, structured_exercise, casual_walk, outdoor_leisure
 \end{Verbatim}

 \paragraph{Distillation Candidate Sets}
 The same configuration file specifies the intention set $\mathcal{I}=\{\texttt{home},\texttt{duty},\texttt{leisure},\texttt{maintenance}\}$, weekday/weekend intention-chain candidates (\path{with_duty_intention_chain} and \path{without_duty_intention_chain}), and the legal mappings \texttt{activity}$\rightarrow$\texttt{intention} and \texttt{activity}$\rightarrow$\texttt{landuse}. These candidate sets define the finite action space used by offline distillation and simulation-time lookup.

\begin{table*}[!htbp]
\centering
\small
\setlength{\tabcolsep}{4pt}
\renewcommand{\arraystretch}{1.05}
 \caption{Reference intention set and allowed activity types used in our instantiation. For each intention $z\in\mathcal{I}$, the teacher scores the predefined candidate set $\mathcal{A}_z$ and we normalize the aggregated scores into a categorical distribution for simulation-time sampling.}
 \label{tab:intention_activity_taxonomy}
 \begin{tabular}{@{}p{0.16\textwidth} p{0.28\textwidth} p{0.52\textwidth}@{}}
 \toprule
 \textbf{Intention $z$} & \textbf{Semantics} & \textbf{Allowed activity types $\mathcal{A}_z$}\\
 \midrule
 \texttt{home} & Stay at residence / rest & \path{sleep_rest}\\
 \texttt{duty} & Obligations (work/school) & \path{work_task, study_class}\\
 \texttt{maintenance} & Daily necessities and errands & \path{daily_shopping, personal_service, medical_care, admin_errand}\\
 \texttt{leisure} & Discretionary activities & \path{solo_meal, social_meal, social_visit, entertainment_activity, structured_exercise, casual_walk, outdoor_leisure}\\
 \bottomrule
 \end{tabular}
 \end{table*}

\FloatBarrier
\subsection{Distillation Prompt Templates}
\label{app:distillation_prompts}

Below are representative prompt templates for offline distillation. Each query type uses a fixed template that includes resident profile fields and outputs structured JSON scores.

\paragraph{Chain Scores Prompt}
\begin{footnotesize}
\begin{verbatim}
Role-play as a resident and score behavior preferences.

Resident: age_bin=<age>, occupation=<occ>
Scenario: typical <day_type>
Candidates: [<chain_1>, <chain_2>, ...]
(H=home, D=duty, L=leisure, M=maintenance)

Task: Score each chain [0-10]. Output JSON only:
{"scores": {"<chain_1>": 5, "<chain_2>": 5}}
\end{verbatim}
\end{footnotesize}

\paragraph{Activity Scores Prompt}
\begin{footnotesize}
\begin{verbatim}
Role-play as a resident and score activity preferences.

Resident: age_bin=<age>, occupation=<occ>
Scenario: pursuing intention='<intention>'
Candidates: [<activity_1>, <activity_2>, ...]

Task: Score each activity [0-10]. Output JSON only:
{"scores": {"<activity_1>": 5, "<activity_2>": 5}}
\end{verbatim}
\end{footnotesize}

Full templates and configuration files are available in the repository at \texttt{data\_prepare/step4\_activity/}.

\FloatBarrier
\subsection{LLM Interface Schema}
\label{app:llm_interface_schema}

This section provides detailed repeatability notes for the LLM-ready interface, including discretization bins, activity--landuse mappings, and missing value handling.

\paragraph{Context Discretization Bins}
Agent context is discretized into coarse bins to enable efficient lookup-table compilation:

 \begin{itemize}
 \item \textbf{Age bins} (3 categories): \texttt{child} (0--17), \texttt{adult} (18--64), \texttt{elderly} (65+)
 \item \textbf{Occupation bins} (9 categories): \texttt{agriculture\_worker}, \texttt{industrial\_worker}, \texttt{service\_worker}, \texttt{office\_worker},\\
 \texttt{professional}, \texttt{public\_sector}, \texttt{self\_employed}, \texttt{non\_employed}, \texttt{college\_student}
 \item \textbf{Day type} (2 categories): \texttt{weekday}, \texttt{weekend}
 \end{itemize}

\paragraph{Activity--Intention Mapping}
Each activity type maps to exactly one intention category:

\begin{footnotesize}
\begin{tabular}{@{}p{0.55\columnwidth}l@{}}
\toprule
\textbf{Activity} & \textbf{Intention} \\
\midrule
\texttt{sleep\_rest} & \texttt{home} \\
\texttt{work\_task, study\_class} & \texttt{duty} \\
\texttt{daily\_shopping, personal\_service, medical\_care, admin\_errand} & \texttt{maint.} \\
\texttt{solo\_meal, social\_meal, social\_visit, entertainment\_activity, structured\_exercise, casual\_walk, outdoor\_leisure} & \texttt{leisure} \\
\bottomrule
\end{tabular}
\end{footnotesize}

\paragraph{Activity--Landuse Mapping}
Each activity type is constrained to specific landuse categories (abbreviations: C=commercial, I=industrial, P=public\_facility, T=transport, O=open\_space, R=residential, A=agriculture, N=nature):

\begin{footnotesize}
\begin{tabular}{@{}ll@{}}
\toprule
\textbf{Activity} & \textbf{Landuse} \\
\midrule
\texttt{sleep\_rest} & R \\
\texttt{work\_task} & C, I, P, T, O, A \\
\texttt{study\_class} & P \\
\texttt{daily\_shopping, personal\_service} & C \\
\texttt{medical\_care, admin\_errand} & P \\
\texttt{solo\_meal} & C, P, T, O \\
\texttt{social\_meal, entertainment} & C, O \\
\texttt{social\_visit} & R, O \\
\texttt{structured\_exercise} & O, P \\
\texttt{casual\_walk} & O, road \\
\texttt{outdoor\_leisure} & O, N \\
\bottomrule
\end{tabular}
\end{footnotesize}

\paragraph{Missing Value Handling}
When agent attributes are incomplete, the following defaults apply:
\begin{itemize}
\item \textbf{Missing occupation}: Mapped to \texttt{non\_employed} bin
\item \textbf{Missing age}: Mapped to \texttt{adult} bin (modal category)
\item \textbf{Missing home location}: Agent excluded from spatial activity generation; flagged as \texttt{no\_location}
\item \textbf{No valid POI for activity}: Fallback to nearest POI of any compatible landuse type; if none available within search radius, activity skipped
\end{itemize}

The complete schema files are available in the repository at \texttt{data\_prepare/step4\_activity/bins\_*.json}.

\FloatBarrier
\subsection{Distillation Setup}
\label{app:distillation_setup_prelim}

We perform offline distillation by repeatedly querying a teacher model under identical discretized context keys $s$ (Section~\ref{sec:distillation_scaling}) and estimating empirical action distributions for each decision query type. Prompt templates used for distillation are listed in Appendix~\ref{app:distillation_prompts}.

\paragraph{Sampling Hyperparameters}
In our reference instantiation, we use the following configuration:
\begin{itemize}
\item \textbf{Repetitions per context key} ($K$): 10 samples per unique $(age\_bin, occupation\_bin, day\_type)$ tuple
\item \textbf{Teacher model}: Gemma~3~27B~\cite{team2025gemma} served locally via Ollama
\item \textbf{Temperature}: 0.7 for score generation (enabling diverse but coherent responses)
\item \textbf{Sampling}: No adaptive sampling; uniform $K$ across all context keys
\end{itemize}

\paragraph{Hardware}
Distillation was performed on a workstation equipped with an RTX~4090 GPU (24GB VRAM), 96GB RAM, and an Intel Core i5-14600K CPU. The teacher model was queried through AgentScope~\cite{gao2024agentscope}.

\paragraph{Unseen Key Handling}
At simulation time, if a context key $s$ was not encountered during distillation (due to rare demographic combinations), we apply a fallback strategy:
\begin{enumerate}
\item \textbf{Coarse-bin fallback}: Map the unseen key to a coarser bin (e.g., specific occupation $\rightarrow$ \texttt{non\_employed})
\item \textbf{Default distribution}: If no matching compiled distribution exists, use a uniform distribution over the candidate action set
\end{enumerate}
In practice, our discretization yields $3 \times 9 \times 2 = 54$ unique context keys for activity preference queries, which are enumerated during offline compilation.

\end{document}